\documentclass[sigconf,screen,nonacm]{acmart}

\AtBeginDocument{
  }

\usepackage{lscape}
\usepackage[shortlabels]{enumitem}

\begin{document}

\title{When Should I Trust This? Designing and Evaluating Factuality Indicators for LLM-based Chatbots}

\author{Greta Warren}
\email{grwa@di.ku.dk}
\orcid{0000-0002-3804-2287}
\affiliation{
  \institution{Department of Computer Science, University of Copenhagen}
  \city{Copenhagen}
  \country{Denmark}
}

\author{Christine M. Lehane}
\email{christine.m.lehane@gmail.com}
\orcid{0000-0001-7872-0890}
\affiliation{
  \institution{Meta Platforms, Inc.}
  \city{Menlo Park}
  \state{California}
  \country{USA}}

\author{Jessica Heal}
\email{jfh@openai.com}
\orcid{0000-0003-4827-6711}
\authornotemark[1]
\thanks{*Work conducted while at Meta}
\affiliation{
    \institution{OpenAI}
    \country{USA}}
\affiliation{
  \institution{Meta Platforms, Inc.}
  \city{Menlo Park}
  \state{California}
  \country{USA}}

\author{Sila Eser}
\email{silaeser.design@gmail.com}
\orcid{0009-0006-9959-2318}
\authornotemark[1]
\affiliation{
  \institution{Meta Platforms, Inc.}
  \city{Menlo Park}
  \state{California}
  \country{USA}}

\author{Denise M. Leto}
\email{denisemleto@gmail.com}
\orcid{0009-0000-4335-7104}
\affiliation{
  \institution{Meta Platforms, Inc.}
  \city{Menlo Park}
  \state{California}
  \country{USA}}

\author{Isabelle Augenstein}
\email{augenstein@di.ku.dk}
\orcid{0000-0003-1562-7909}
\affiliation{
  \institution{Department of Computer Science, University of Copenhagen}
  \city{Copenhagen}
  \country{Denmark}
}

\renewcommand{\shortauthors}{Warren et al.}

\begin{abstract}
Generative AI systems increasingly mediate how people seek information, yet communicating AI factuality to laypeople remains challenging. Through a mixed-methods investigation involving interviews (N=8), a survey (N=1,000), and an experiment (N=5,151), we examine how people evaluate AI-generated information for gardening, mathematical, and medical queries, and how factuality indicators (confidence and sources) shape perceptions. Our interviews and survey show people selectively verify AI responses using external and AI-provided evidence, moderated by contextual factors. Although confidence-indicators affected trust and accuracy perceptions, people hold diverse, ambiguous interpretations of AI confidence. Sources were viewed as crucial, yet users raised concerns about source reliability and verification effort. Our experiment showed that confidence and source-based indicators influenced initial perceptions, but effects disappeared once participants learned the response's correctness. Medical queries elicited skepticism regardless of indicator. Findings motivate helping users interpret model confidence and integrating source information into the interface to support calibrated trust in AI-supported information-seeking.

\end{abstract}




\maketitle

\section{Introduction}
Large language models (LLMs) are increasingly used as tools for information-seeking \cite{simon2025generative,ceron2026wildseekevaluatinglanguagemodels} due to their growing prominence in online search and social media platforms \cite{grossman2026AIdisruptssearch,kirsten-etal-2026-characterizing}, and conversational interfaces.
Yet the fluency of LLM-generated responses does not guarantee their factuality: models can produce plausible-sounding but inaccurate, hallucinated text that is not reliably grounded in any verifiable source \cite{pauli-etal-2025-measuring,augenstein2024factualityLLMs,ji2023hallucination}, which can lead people to follow misleading advice.

Prior work has explored two main approaches to helping users assess the factuality of AI-generated responses: communicating the model's own confidence \cite{kim2024LLMuncertainty,prabhudesai2023uncertainty,bhatt2021uncertaintytransparency,Vasconcelos2025generationprob}, and providing sources or citations that allow a claim to be independently checked \cite{kim2025fostering,warren2026evidence}.
While these approaches are promising, findings suggest that their impacts depend heavily on how the information is presented, and it is unclear how such indicators should be designed.

As AI chatbots take on a larger role in how people access and act on information, understanding how everyday users evaluate and make decisions about when to trust an AI-generated response is essential to designing systems.
However, large-scale studies assessing people's verification behaviours and attitudes are scarce, making it difficult to design effective factuality indicators for everyday chatbot usage.

Furthermore, most empirical studies have been focused on specific tasks, such as fact-checking \cite{warren2026evidence}, however, people increasingly use AI chatbots for everyday personal questions related to hobbies, work, and health. Comparatively little is known about how everyday LLM chatbot users perceive and interpret confidence- and source-based indicators.

To address these gaps, we posed the following research questions:

\textbf{RQ1:} How should factuality indicators be designed to help people assess AI factuality?

\textbf{RQ2:} How do laypeople assess the factuality of information from AI chatbots?

\textbf{RQ3:} How do factuality indicators affect perceptions of AI chatbots?

We address these questions through a mixed-methods approach comprising three complementary studies.
First, we conducted in-depth interviews with eight AI users to surface how they perceive AI confidence, sources, and verification behaviours (\textbf{RQ1}), as well as their reactions to speculative design concepts for factuality indicators.
Second, we conducted a nationally representative survey of 1,000 U.S. AI users to characterise how, when, and why people currently verify information from AI chatbots, and what additional information and support might be useful (\textbf{RQ2}). Third, we conducted a large-scale between-participants experiment (N=5,151)
that systematically varied the type of factuality indicator shown (confidence-labels, confidence-percentages, confidence-bubbles, source-labels, sources and confidence language, and no indicator), the topic of the query (gardening, mathematics, medical advice),
and the correctness of the response,
to evaluate how different factuality indicators influence perceived accuracy, information sufficiency, trust, and sentiment, both before and after participants learned whether the response was actually correct (\textbf{RQ3}).

Our interviews revealed that while users value confidence indicators, they hold diverse and often conflicting interpretations of what confidence means. Sources were perceived as highly important, given that they provide evidence that can be independently evaluated, although users also described the effort and other limitations associated with source verification.
Our survey showed that most people verify AI responses at least some of the time, but they primarily do so through external information seeking and source inspection rather than through AI-provided confidence information.
Finally, our experiment found that both confidence and source indicators have a small effect on people's perceptions of an AI chatbot before a response's correctness is known, yet this effect dissipates once the correct response is revealed. We further find that contextual factors, particularly the stakes associated with a task, influence the effectiveness of factuality indicators.

\section{Related Work}

\subsection{Assessing AI factuality}
Although LLM chatbots can produce fluent, convincing outputs, they often also produce false information that can mislead their users and others \cite{ji2023hallucination,augenstein2024factualityLLMs}.
This can lead people to erroneously trust or follow AI advice, often termed overreliance \cite{ibrahim2025measuringmitigatingoverreliancenecessary}, which can lead to inaccurate decisions, skill erosion, and other potential harms \cite{miller2023explainable,bucinca2025decisionskills}.
The superficially plausible and persuasive nature of LLM outputs makes it important to be able to distinguish factual information from false information, and so recent work in human-AI interaction has explored methods of supporting AI users to assess the factuality (that is, a response that is truthful, or based on fact \cite{augenstein2024factualityLLMs,ji2023hallucination,do2025highlight}), of an AI response, so that they can decide whether or not the response is trustworthy \cite{lee2004trust}.

Recent work has shown that factuality indicators, that is, design features that provide the user with additional information to assist with evaluating the factuality of a response, can be useful supports and help people to appropriately calibrate their reliance on AI advice.
For example,
Do et al. \cite{do2025highlight} show that incorporating in-line highlighting in LLM responses based on factuality levels made it easier for users to validate the accuracy of the LLM's response and increased their trust, compared to a baseline where no style was applied.
Another study found that highlighting uncertain parts of the model output did reduce overreliance, but also made people more likely to reject correct AI responses \cite{bo2025relianceinterventions}, while other approaches have shown that removing factually incorrect information from responses can enhance user trust \cite{DoGeyer2025hideorhighlight}.
Although various methods of signalling the factuality of AI responses have been studied, it remains unclear how these various methods are understood and used by people in everyday interactions with AI chatbots across various contexts.
In this paper, we study two prominent categories of factuality indicators for LLM-based chatbots -- confidence and source indicators -- in order to understand how they impact people's perceptions of accuracy and trust in AI, and how they can be designed to more effectively support people in AI-supported information-seeking.

\subsection{Effects of communicating confidence and uncertainty}

Various methods of communicating AI uncertainty to human users have been proposed with the aim of allowing people to better calibrate their trust and reliance in human-AI decision-making \cite{bhatt2021uncertaintytransparency,chiaburu2024uncertainty}.
For example, pre-LLM work has found that presenting numerical confidence (e.g., percentages) to users can improve task performance \cite{bansal2021teamperformance} and trust calibration \cite{zhang2020confidence}.
Confidence estimates and explanations have been found to have complementary effects on people's accuracy in predicting a model's output \cite{papantonis2025not}, while other work has found that automatically highlighting snippets of code that are likely to be edited helped programmers to complete their tasks faster and were subjectively preferred \cite{Vasconcelos2025generationprob}.
Visualising AI uncertainty has also been proposed as a method of getting people to slow down and think more deliberatively \cite{prabhudesai2023uncertainty}.
Although these methods facilitate precise, granular uncertainty quantification, they can be challenging for people to interpret \cite{hofman2020uncertainty} and put into action \cite{warren2025explainablefactchecking}.
Uncertainty communicated through natural language may be more intuitively understood, though people also differ in how they interpret them (e.g., individual differences for what people consider ``very confident'' tend to be large) \cite{steyvers2024calibrationgapmodelhuman,belem-etal-2024-perceptions,wallsten1993preferences,zimmer1983verbal}.
Recent work in LLM-specific settings has found that how an LLM verbalises uncertainty in its own generated text (e.g., by prepending statements like ``I'm not sure, but...'') shapes both reliance and trust \cite{kim2024LLMuncertainty}.
However, the same study also found that people's perceptions and behaviours are highly sensitive to how such verbalised uncertainty is phrased.
It remains unclear how different styles of confidence indicators shape people's perceptions of AI chatbots, and how they should be designed to support factuality assessment more effectively.
In this study, we address this by examining how people interpret various sorts of confidence indicators (numerical percentages, categorical labels, and natural language) and systematically comparing how they impact perceptions and trust in AI responses.

\subsection{Effects of providing sources, evidence and citations}
An intersecting line of research has examined how providing sources in various forms alongside a generated response can support people to assess AI factuality.
Presenting sources in the form of citations or links has become a common design pattern in commercial LLMs (e.g., ChatGPT, Claude, Gemini), however these approaches are not always reliable: they can misrepresent the original source (e.g., \cite{liu-etal-2023-evaluating}) as well as fabricate non-existent sources (e.g., \cite{alkaissi2023artificial}).
Kim et al. \cite{kim2025fostering} found that providing sources alongside AI responses made people more confident in their responses in an AI-supported information-seeking task and appeared to increase the credibility of the LLM response. They also found that verifying the sources increased people's performance in determining when they could or could not rely on the AI system's answer, though it did not eliminate overreliance entirely.
However, some studies have shown that people often do not verify sources when provided as clickable links to external information \cite{kim2024LLMuncertainty}, while other studies indicated that presenting evidence from sources in a more integrated way (e.g., dropdown buttons) can improve people's engagement with sources when evaluating whether a fact-checking claim is true or false \cite{warren2026evidence}.
This evidence suggests that the way in which sources are presented to users plays a significant role in their verification behaviour. However, the best approaches to designing source indicators to support factuality assessment are unclear.
Furthermore, much of the existing literature focuses on tasks in which participants must evaluate AI responses focused on general knowledge or fact-checking question-answering, rather than on everyday LLM chatbot users who are concerned with specific personal queries.
In the current paper, we examine the role of sources in people's verification behaviour, how they determine source credibility, and how providing sources alongside AI responses impacts people's perceived accuracy, trust, information sufficiency and impressions of an AI chatbot, across a variety of topics, ranging from benign everyday queries to high stakes medical ones.

\section{RQ1: How should factuality indicators be designed to help people assess AI factuality?}

To investigate and develop an initial understanding of how laypeople perceive and reason about AI factuality cues, including confidence indicators and source-based verification features, we conducted an exploratory qualitative study with AI chatbot users.

\subsection{Method}
We conducted semi-structured interviews to enable participants to articulate existing mental models of accuracy, trust, and verification while providing detailed feedback on a set of speculative design concepts.
Interviews were conducted in March 2025.

\subsubsection{Participants.}
Eight participants from the United States were recruited to participate in one-on-one virtual interviews (see Table \ref{tab:interview_ppts} for an overview of demographics).
All participants reported experience using generative AI tools either weekly or daily.

\begin{table}[h]
    \centering
    \begin{tabular}{cccc}
\toprule
\textbf{Participant ID} & \textbf{Gender} & \textbf{Age} & \textbf{AI Usage} \\
\midrule
P1 & Female & 29 & Weekly \\
P2 & Female & 23 & Daily \\
P3 & Female & 23 & Weekly \\
P4 & Male & 17 & Daily \\
P5 & Male & 25 & Daily \\
P6 & Male & 27 & Daily \\
P7 & Male & 17 & Weekly \\
P8 & Female & 16 & Weekly \\
\bottomrule
\end{tabular}
\caption{Interview participant demographics}
    \label{tab:interview_ppts}
\end{table}

\subsubsection{Interview Protocol.}
Each participant completed a 60-minute remote interview. Sessions were audio-recorded.
Interviews began with a warm-up discussion about how participants used AI, and how much they relied on them.
The main interview focused on five topics:
\begin{enumerate}
    \item \textit{Current behaviours and expectations around information accuracy:} Participants were asked about how they assessed the accuracy of information from (i) people they knew, (ii) news and social media, and (iii) AI tools. Questions focused on what evidence they expected to be provided, how they determined source credibility, and how sources (could) present indicators of credibility or accuracy.
    \item \textit{Communicating AI accuracy:} Participants were prompted to imagine how AI confidence could be communicated in a useful way, and react to design stimuli illustrating AI confidence indicators such as labels and percentages. Participants were also asked what they considered acceptable levels of confidence, and what they interpreted such confidence labels to be based on.
    \item \textit{Using confidence indicators:} Questions focused on how AI factuality indicators might be useful to them, and what additional information and explanations they might require. Participants were then asked to react to further design stimuli showcasing additional information about sources or confidence and discuss their utility and preferences.
    \item \textit{Citations and sources:} Participants were asked about their experiences with AI-provided sources and their verification behaviours. They were shown design stimuli showing examples of how source information could be provided (e.g., pop-ups or inline with responses) and asked to evaluate them and imagine how they could be used and improved.
    \item \textit{Contextual factors:} Participants were asked to imagine how they would react in certain situations, such as if an AI chatbot's response was labelled with 'low accuracy', or if a confidence indicator was misaligned with the actual accuracy, as well as about how specific topics or situations might impact how important or useful AI factuality indicators would be.
\end{enumerate}

Finally, participants were asked for any further thoughts or comments they had.
Throughout the interview, participants were asked follow-up questions, and encouraged to explain their reasoning, describe their expectations about what an AI chatbot should do, and compare alternatives.

\subsubsection{Design Stimuli.}
\begin{figure*}
    \centering
    \includegraphics[width=\linewidth]{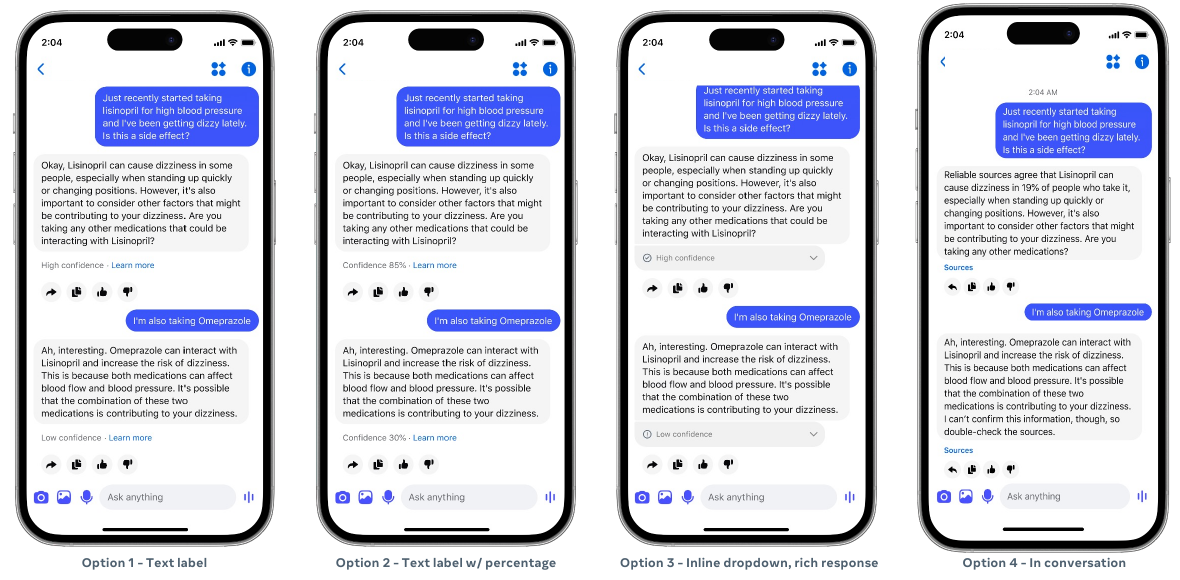}
    \caption{Designs of confidence indicators shown to interview participants (Design Stimulus A)}
    \label{fig:interview_stimA}
\end{figure*}

We created a set of design concept images which were presented during the interviews to elicit participant reactions, expectations, and explore other possible approaches for communicating AI factuality.
We designed the images to show a conversation with an AI chatbot on a mobile device, to resemble a scenario in which people use AI to seek information for a medical query.
We chose this medical context to prompt people to reflect on how to assess AI factuality for a  high-stakes topic, similar to previous work (e.g., \cite{kim2024LLMuncertainty}).
The content consisted of a two-turn interaction with the chatbot, in which the user first makes a query about side-effects of a medication, and the chatbot response probes for more information. The response is accompanied by a high confidence indicator.
Then, the user responds to the follow-up question, and the chatbot provides a response, this time with a low confidence indicator.
All participants were shown the same design concepts, but the order of different options for the same concept was randomised between participants (e.g., one participant saw the options in Fig \ref{fig:interview_stimA} in the order: Option 3, Option 1, Option 2, Option 4).

Design Stimulus A (see Fig. \ref{fig:interview_stimA}) presented four different versions of AI chatbot confidence indicators: (i) \textit{Confidence-labels} (high/low), (ii) \textit{Confidence-percentage} (85\%/30\%), (iii) \textit{Confidence-labels with inline dropdown} (check mark / warning symbol and dropdown arrow), and (iv) \textit{Confidence-aligned language} (``reliable sources agree'' vs ``I can't confirm this information''). We chose these four indicator types to exemplify different approaches: confidence labels for categorical assessment, confidence percentages for quantitative, more granular information, dropdown for providing additional detail on demand, and confidence-aligned language for a more natural conversational indicator as opposed to specific interface details.

Design Stimulus B (see Fig. \ref{fig:designstimulusB}) showed percentages underneath the AI responses with three different candidate labels: confidence, reliability, and certainty, to understand how different linguistic framings shaped people's interpretations of AI-generated information and gauge preferences. Each design was otherwise identical.

Design Stimulus C (see Fig. \ref{fig:designstimulusC}) showed two  ways users could learn why a response received a particular confidence rating. The first example, an explanatory bottom-sheet, showed a pop-up that appeared after interacting with the confidence indicator and explained that the confidence rating was based on the quality of the sources.
The second example showed source information in the form of an expandable inline dropdown, which users could read without navigating away from the chat interface.

Design Stimulus D (see Fig. \ref{fig:designstimulusD}) showed three examples of how AI chatbots could provide sources and citations.
In the first example, clicking on an in-text citation triggered a pop-up panel containing excerpts from the cited webpage, and the URL to the specific page used to generate the answer.
In the second example, specific text was highlighted within the response, and clicking on this text brought up a pop-up panel also containing a URL and excerpts from a relevant webpage.
In the third example, a bubble underneath the main response which read `Researched' could be clicked which would dropdown to reveal the number of sources consulted and an overview of the source.

\subsubsection{Data analysis.}
We followed an inductive thematic analysis approach. The first author read the interview transcripts to familiarise themselves with the data. They then conducted the primary analysis using a bottom-up coding process. After coding four interviews, the first author iteratively compared codes across participants and grouped related codes into broader conceptual categories (e.g., `verification cues' or `high-stakes contexts'), before coding the next 1-2 interviews, repeating the process until all transcripts had been analysed.
The broader research team then discussed the emerging concepts, reviewing and refining the category structure and iterating over larger themes
until consensus was reached regarding the key final themes.

\subsection{Findings}
Our analysis surfaced insights into how participants used and interpreted confidence and source indicators, how these features impacted their verification behaviour, as well as how their use and perceptions of these indicators was shaped by contextual factors such as topic, stakes, and complexity.

\subsubsection{Confidence indicators}

Participants agreed that knowing how confident a chatbot was in its response would be beneficial.
Specifically, participants mentioned using confidence indicators as cues to determine when they could rely on an answer and when additional verification might be required.
For example,
several participants described confidence indicators as a cue for deciding whether to perform further fact-checking or search for information independently.
One participant explained that confidence could help determine ``{how much research I should put in myself}'' (P7).
Confidence was therefore viewed by some people less as a direct measure of truth and more as a guide for allocating verification effort.

Participants also frequently described confidence indicators as useful for aligning their expectations with model accuracy:
``it's good to know that you can recognize... that the system can recognize whether it's fully accurate or not and if it's very confident or not.'' (P3)
Participants often felt that low-confidence indicators would prepare them for possible inaccuracies and reduce disappointment if an answer proved incorrect. As P5 explained:
``at least I understood that it was low confidence and it was incorrect.''
Similarly, P1 interpreted low-confidence signalling as an act of transparency: ``They're like, you can't trust what we say, but we want to let you at least know we're trying.''
In contrast, participants reacted strongly to imagined scenarios where confidence indicators were incorrect. While low confidence paired with a correct answer was generally viewed as confusing but acceptable, high confidence paired with an incorrect answer was perceived as a serious issue that eroded participants' trust and would discourage future use of the system. As P7 stated:
``I'll be very mad. And I would probably not use AI again.''

\paragraph{Communicating confidence}
Although participants unanimously agreed on the importance of knowing how confident a chatbot was in its responses, many differing definitions and interpretations of confidence emerged during the interviews.
Some participants described percentage confidence as the probability that the AI's response was correct, while others conceptualised it as the  proportion of the response that was accurate, or the proportion of the response that was based on the source that it cited. As P7 explained: ``So it's kind of giving off the idea, well, to me that it's 100\% accurate to what the source said.''
One participant expressed the belief that high confidence meant that the AI system had already performed verification on their behalf:
``Because if you tell me like, you're very confident your answers, I know, okay, you double-checked, you verified. So I wouldn't need to do that'' (P3).
Interestingly, several participants shifted between different interpretations of confidence during the same interview, i.e., their descriptions of how they interpreted AI confidence suggested multiple, conflicting mental models of what this confidence represented. These findings suggest that confidence indicators may be susceptible to multiple competing mental models, creating ambiguity around their intended meaning.

Participants also frequently highlighted that they expected explanations describing how confidence scores were generated, and what information they were based on.
P7 suggested making uncertainty more concrete by highlighting certain parts of the AI response that were less confident:
``a disclaimer at the bottom saying like a percent or sentence saying I feel this is [...] 100 confident, 50 confident, or highlights or bolds the parts and saying bolded parts, please fact-check.''

\paragraph{Presentation and Context}
Most participants preferred numerical confidence percentages over categorical labels such as `high' or `low' confidence. Numerical values were perceived as more `exact' and precise.
However, this preference was not universal. Some participants argued that percentages were inherently vague because they did not explain how the values had been calculated or what exactly they represented.

Participants generally agreed that AI confidence was particularly important for high stakes information, such as ``health or career related finances'' (P1).
Several participants also distinguished between simple and complex queries. For straightforward questions with clear answers, categorical indicators such as `high confidence' were often considered sufficient. However, for nuanced or consequential topics, participants preferred numerical confidence estimates that provided additional detail. For example, P7 stated:
``since it's a drug and relates to health, I would prefer like I would feel more safe and trusting if it said the percent. Because you know [...] if it were my health, I wouldn't just want to be high confidence, I would want it to be 100\%'' (P7)

Similarly, the same participant suggested that percentages become more useful as questions become more complex:
``I think it really depends what I'm asking for, how much research goes into the question. If it's just a one sided question, then yeah, I would like high confidence. But since this is an in-depth question, there are a lot of aspects, I do like the percent level.''

Although participants valued confidence indicators, many argued that confidence information alone was insufficient. Instead, participants frequently requested accompanying sources that would allow them to understand why a response received a particular confidence score and independently assess its credibility. Confidence indicators were therefore often interpreted as a signal to inspect supporting evidence rather than as evidence in themselves.

\subsubsection{Source indicators}
Participants generally viewed sources as useful, and in some cases, crucial elements to accompany AI responses.
Sources served several functions: they acted as proof for claims, provided tangible evidence that could be inspected directly, and helped users determine whether information was trustworthy.
As P5 explained: ``I just like having sources and like just showing proof of like what it's mentioning.''
Similarly, P1 described the importance of having information that users could explore further for themselves: ``Having an article, something to click on, something to learn more information about, something to pinpoint more.''
Several participants also interpreted sources as evidence that the AI system had actively gathered supporting information. As P7 noted: ``But with the link it makes it seem like they actually went in and did research to formulate this answer.''

\paragraph{Source credibility}
Participants referred to a variety of source types when describing credibility, including
mainstream news organisations, government websites, and academic research papers.
Credibility was also linked to reputation and prior experience. P1 described relying on collective judgments, and reputation built over time to assess source quality: ``From [...] people that used the sources before, people that know what it's about [...] Knowing people used it before, knowing that it's provided accuracy for years.''
However, participants varied in how many sources they considered sufficient to accompany a response.
For example, some participants took the view that ``the more sources the better'' or that ``at least 10 good sources'' might be appropriate.
Others believed that ``one source is enough'', provided it was credible.
P5 emphasised the importance of source diversity, suggesting that AI responses should include: ``three different places, three different sources.''

\paragraph{Limitations of sources}
Despite placing high importance on the role of sources in assessing AI factuality, participants also identified several concerns about providing sources alongside AI responses.

First, checking sources can require considerable effort. While sources provide opportunities for verification, many participants acknowledged that they may not always take the time required to inspect them:
``Yeah, I'll click on it, but then let's say I'm in a rush. I probably would not want to because then I would have to do extra work to like, probably [check] the information myself.'' (P3)

Second, participants expressed concerns about retrieval accuracy.
Some participants had experienced doubts regarding whether an AI system had correctly interpreted their queries and selected relevant information. P1 expressed uncertainty about whether the system would retrieve the appropriate content: ``If they interpreted my answer right, if they look up for the right information.''
P7 observed that AI inaccuracy, despite there being correct answers available online, was particularly frustrating: ``There are accurate answers all over the Internet ... and it's giving me inaccurate responses, then I'd prefer just to go to the source itself.''

Third, participants were sceptical regarding  whether AI systems could reliably provide genuine sources. Several participants described past experiences where AI systems generated fabricated or inaccessible source links. As P7 explained: ``Because in my experience I got sources and links that led to fake websites or websites that didn't exist.''

\paragraph{Context and presentation}
Participants described how task context influenced how likely they were to check sources. Academic work and healthcare questions were frequently identified as situations where sources would be particularly beneficial:
``People that's like maybe doing a project or something for college or, you know, they want definitely accurate information.'' (P1)

Participants also had clear preferences regarding source presentation.
Participants tended to prefer dropdown designs over pop-up ones, mentioning that sources should be easy to locate, directly accessible, and connected to specific claims within the AI response.
Participants also preferred designs that presented excerpts or pull quotes from source material because these reduced the effort required to verify information and helped identify potential hallucinations: ``it's saying exactly which part was taken from the source'' (P7). Rather than simply providing links, participants wanted transparency about exactly which information was being used to support a claim.

The analysis of the semi-structured interviews provided rich insights into how laypeople use AI chatbots for information-seeking in their everyday lives.
Building on these qualitative findings, in the next section, we developed a survey and collected data from a large U.S. representative sample, in order to gain broader understanding of how people evaluate AI factuality.

\section{RQ2: How do laypeople assess the factuality of information from LLM chatbots?}

We conducted an observational survey with the aim of understanding people's AI verification behaviours.

\subsection{Method}

\subsubsection{Participants}
The sample consisted of 1000 adults recruited through YouGov's online research panel in the United States (48.1\% female, 51.9\% male), aged 18-86 ($M_{age}$ = 46.26, $SD_{age}$ = 16.24).

We recruited the sample through YouGov's opt-in panel \cite{yougovMethodology}.
The sample was stratified based on gender, age, race and education, based on the 2023 Current Population Survey (CPS) to be representative of US adults.
Participants were pre-screened according to their Internet and AI use.
To qualify, respondents were required to report using at least one of nine popular AI systems (ChatGPT, Gemini, Copilot, Meta AI, Perplexity, Claude, DeepSeek, Grok, or Apple Intelligence) at least once per month. Respondents who reported less frequent use were screened out.
1,586 respondents were screened out due to not meeting this criteria. A further 158 respondents did not complete the survey, while 334 were omitted following quality control checks
and 225 were omitted due to already meeting demographic quotas.
Data collection took place in May 2025.

\subsubsection{Survey Design}
We administered a web-based survey examining how people assess the accuracy and trustworthiness of generative AI responses (see Table \ref{tab:survey_questions} for survey items).
We collected responses about people's verification behaviours when using AI chatbots, such as how frequently they verified AI information, what steps (if any) they took to do so, and why they did so. We also asked people about information that AI chatbots could provide to help them evaluate their reliability.
We also collected respondents' demographic information (e.g., age, gender).
The survey took approximately 5 minutes to complete.

\begin{table*}[]
\begin{tabular}{ll}
\hline
\textbf{Survey Question} &
  \textbf{Response Options} \\ \hline
\begin{tabular}[t]{@{}l@{}}\textbf{AI Use Cases:} Which of the following, if any,\\ have you used generative AI chatbots for in the \\ past 30 days? Select all that apply.\end{tabular} &
  \begin{tabular}[t]{@{}l@{}}(a) Coding or programming; (b) Homework or studying;\\ (c) Exploring what AI can or will do; (d) Chatting, joking, \\ debating, or creating a story; (e) Research or recommendations; \\ (f) Productivity, planning, or organization; (g) Role-playing or \\ playing games; (h) Language or translation help (i) Writing or \\ editing; (j) Creating, editing, captioning, or identifying images;\\ (k) Personal advice or support; (l) Ideas or inspiration;\\ (m) Other (please specify)\end{tabular} \\
  \hline
\begin{tabular}[c]{@{}l@{}}\textbf{AI Perceptions:} Thinking about your \\ interactions  with generative AI chatbots, \\ please rate the degree to which you find AI-\\ generated responses to be:\end{tabular} &
  (a) Believable; (b) Accurate; (c) Trustworthy; (d) Biased \\
  \hline
\begin{tabular}[c]{@{}l@{}}\textbf{Accuracy Evaluation Confidence: } How \\  confident are you in your ability to evaluate \\ the accuracy of the responses you get from\\   generative AI chatbots?\end{tabular} &
  \begin{tabular}[c]{@{}l@{}}(1) Not at all confident (2) Somewhat confident (3) Moderately \\ confident (4) Quite confident (5) Extremely confident\end{tabular} \\
  \hline
\begin{tabular}[c]{@{}l@{}}\textbf{AI Inaccuracy:} In the past 30 days, how often\\  have you received responses from generative \\  AI chatbots that were false or inaccurate?\\ \end{tabular} &
  (1) Never, (2) Rarely, (3) Sometimes, (4) Quite often, (5) Very often \\
  \hline
\begin{tabular}[t]{@{}l@{}}\textbf{Verification Behaviours:} When using \\generative AI chatbots, how often do you\\  take the following steps to verify the\\  accuracy of a response?\end{tabular} &
  \begin{tabular}[t]{@{}l@{}}(a) Look up additional information online to confirm the \\ response; (b) Review the sources provided by the AI;\\ (c) Compare responses from multiple AI chatbots; (d) Check \\ what others have said about the topic online; (e) Discuss the \\ topic with a friend, colleague, or family member; (f) Discuss \\ the topic with someone who has expertise (g) Request a \\ confidence level or rating from the AI.\end{tabular} \\
  \hline
\begin{tabular}[t]{@{}l@{}}\textbf{AI Confidence:} Consider a scenario where \\ you're using an AI chatbot to research the\\  topics listed below. For each topic, rate how\\  important it would be for  you to know when \\  the AI chatbot is not  confident in its response.\end{tabular} &
  \begin{tabular}[t]{@{}l@{}}(a) Everyday information needs (e.g., recipe ideas, trip planning);\\ (b) Research tasks such as school or work assignments; (c) Financial \\ planning and decision-making; (d) Legal matters and advice; \\ (e) Health or medical advice; (f) Civic engagement and voting \\ information; (g) Emotional support and counseling; (h) Relationship \\ and family advice; (i) Career development guidance; \\ (j) Roleplaying, entertainment or games\end{tabular} \\
  \hline
\begin{tabular}[t]{@{}l@{}}\textbf{Verification Motivations: } What motivates \\you to double-check AI- generated information? \\ Select all that apply.\end{tabular} &
  \begin{tabular}[t]{@{}l@{}}(a) I want to test how accurate it is; (b) I don't trust AI very much;\\ (c) The subject matter is important or high stakes; (d) The information \\ provided by the AI seems suspect or unreliable; (e) I want to ensure \\ accuracy for my own information; (f) I want to ensure accuracy before \\ I share the information with others; (g) I'm concerned about potential \\ biases in the AI's response; (h) Other, please explain; (i) I don't \\ double-check AI-generated information\end{tabular} \\
  \hline
\begin{tabular}[c]{@{}l@{}}\textbf{Information Sufficiency:} To what extent do \\ you think generative AI chatbots provide \\sufficient information to help you evaluate \\ the trustworthiness of a response?\end{tabular} &
  (1) Not at all, (2) A little, (3) Somewhat, (4) Quite a bit, (5) Very much \\
  \hline
\begin{tabular}[t]{@{}l@{}}\textbf{Verification Attitudes:} Please indicate how \\strongly you agree or  disagree with the \\following statements:\end{tabular} &
  \begin{tabular}[t]{@{}l@{}}(a) Fact-checking AI responses takes too much time; (b) I'm \\ uncertain about how to verify the accuracy of AI responses;\\ (c) I assume that AI responses are accurate unless proven \\ otherwise; (d) I don't fact-check AI responses unless the \\ information seems obviously false; (e) I often double-check \\ AI responses for accuracy;\end{tabular} \\ \hline
\end{tabular}
\caption{Survey questions and response options}
\label{tab:survey_questions}
\end{table*}

\subsection{Survey Results}

\subsubsection{AI chatbot usage contexts}
\begin{figure*}
    \centering
    \includegraphics[width=\linewidth]{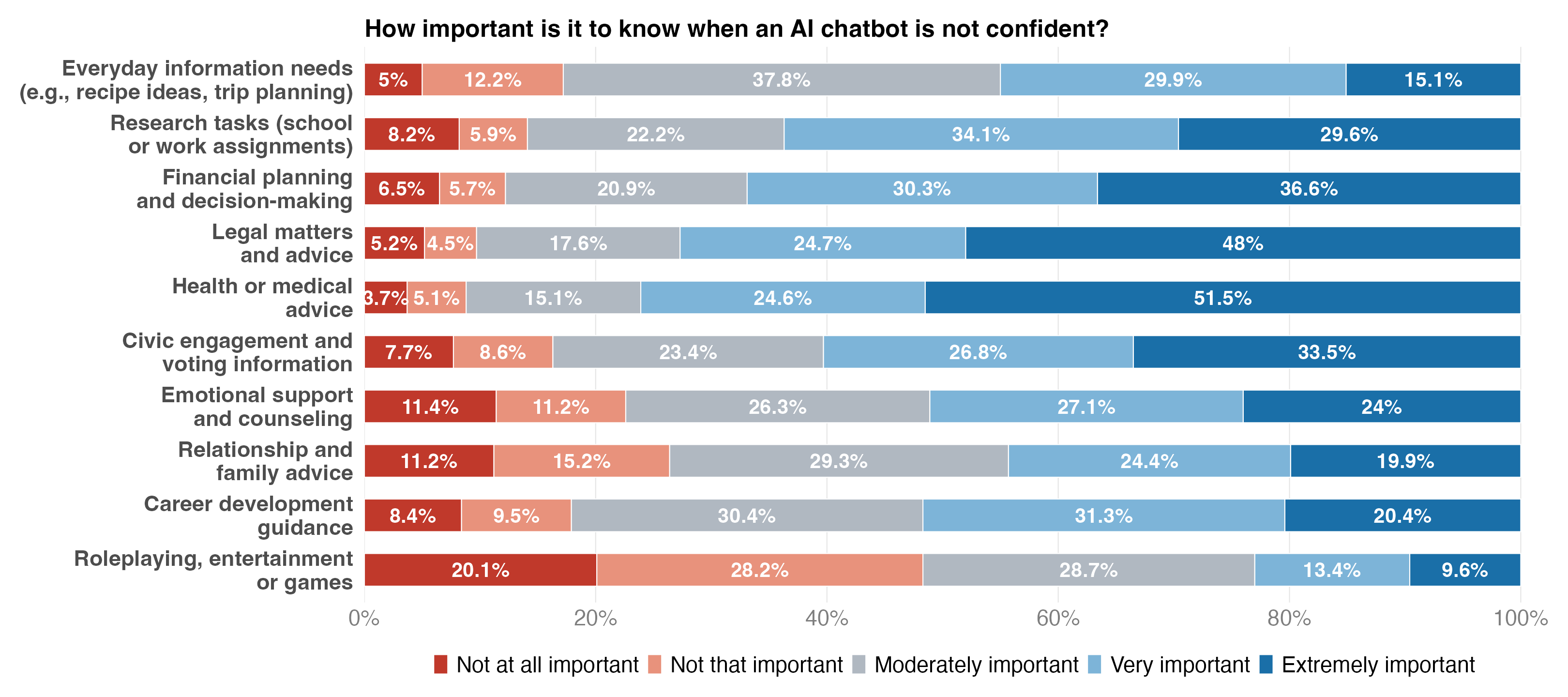}
    \caption{AI confidence use cases}
    \label{fig:survey:usecases}
\end{figure*}

People reported using AI for a diverse range of tasks.
The three most common were (1) research \& recommendations (49.9\%), (2) writing \& editing (38.6\%), and (3) ideas \& inspiration (35.2\%), while studying (15.5\%), coding (8.9\%) and role-playing (8.8\%) were less common use cases (see Fig. \ref{fig:app:survey_usecases} in \ref{app:rq1:results} for details).
People's responses indicated that knowing AI confidence would be beneficial across a broad range of contexts (see Fig. \ref{fig:survey:usecases}).
Importance ratings were highest for health and medical advice (79\%), legal matters and advice (77\%), and financial planning and decision-making (72\%), and lowest for everyday information needs (59\%), relationship and family advice (58\%) and roleplaying, entertainment or games (42\%).

\subsubsection{Perceptions and experiences of AI chatbots}
\begin{figure}
    \centering
    \includegraphics[width=\linewidth]{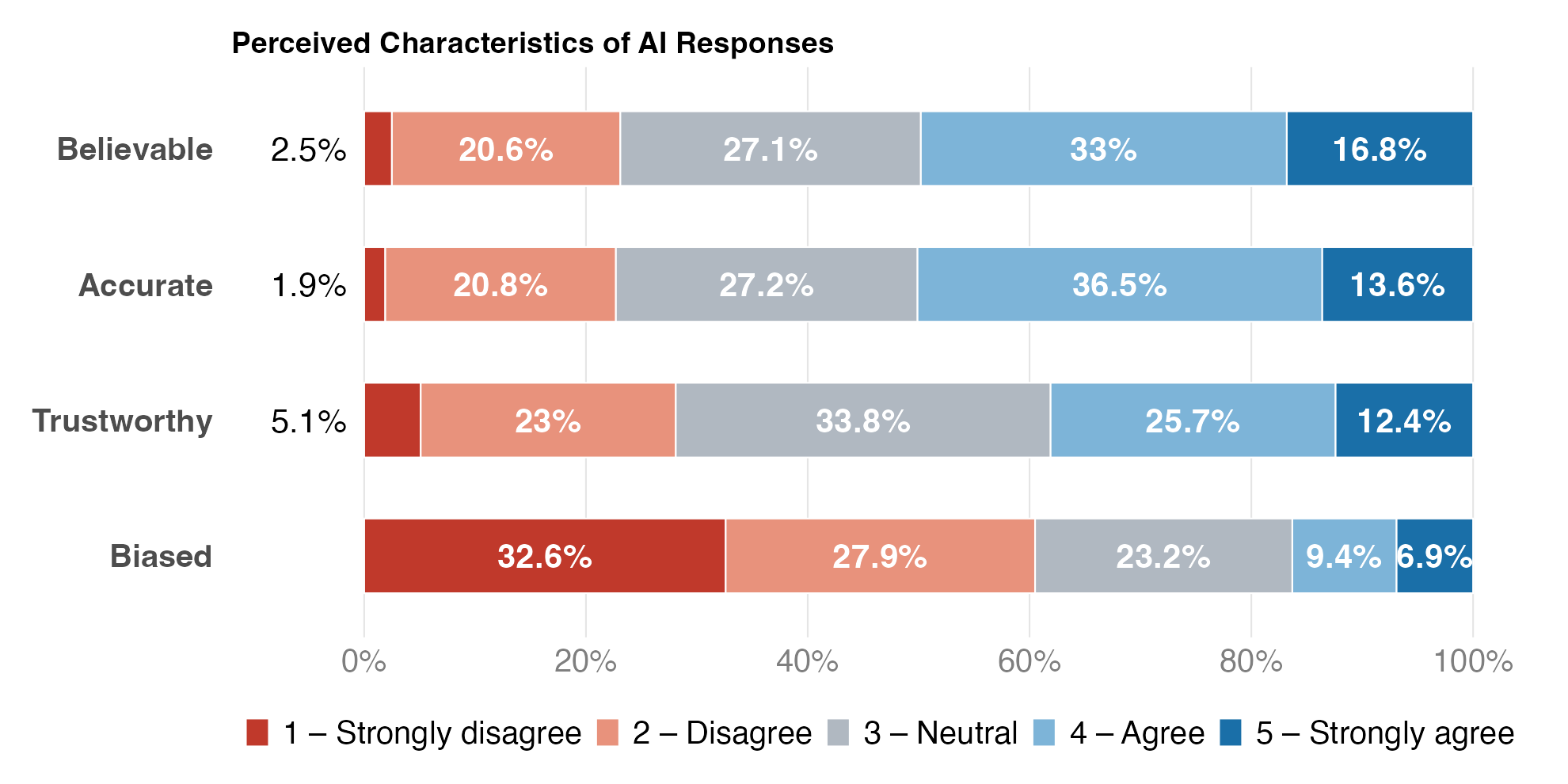}
    \caption{Perceived characteristics of AI responses}
    \label{fig:survey:AI_characteristics}
\end{figure}

Overall, people reported mixed, slightly positively skewed experiences and perceptions of AI reliability (see Fig. \ref{fig:survey:AI_characteristics}).
People judged that AI generated responses that they had previously encountered were moderately believable ($M$ = 3.41, $SD$ = 1.07), accurate ($M = 3.39, SD = 1.02$), and trustworthy ($M$ = 3.14, $SD$ = 1.08), and somewhat biased ($M$ = 2.30, $SD$ = 1.21).
Most people (81\%) also reported coming across inaccurate information from AI chatbots at some point in the previous 30 days
(6\% Very often; 13\% Quite often; 34\% Sometimes; 27\% Rarely; 19\% Never).
Overall, people reported moderate levels of confidence in their ability to evaluate AI accuracy: 95.4\% said that they were at least somewhat confident in this capability (13.8\% Extremely confident; 30.9\% Quite confident; 34.5\% Moderately confident; 16.2\% Somewhat confident; 4.6\% Not at all confident).
People reported that current AI chatbots did provide some degree of information to evaluate AI trustworthiness, but only 10.1\% believed that they did to a very great extent (30.9\% Quite a bit; 43.3\% Somewhat; 11.7\% A little; 4\% Not at all).

\subsubsection{Verification behaviours, motivations, and attitudes}

\begin{figure}[h]
    \centering
    \includegraphics[width=\linewidth]{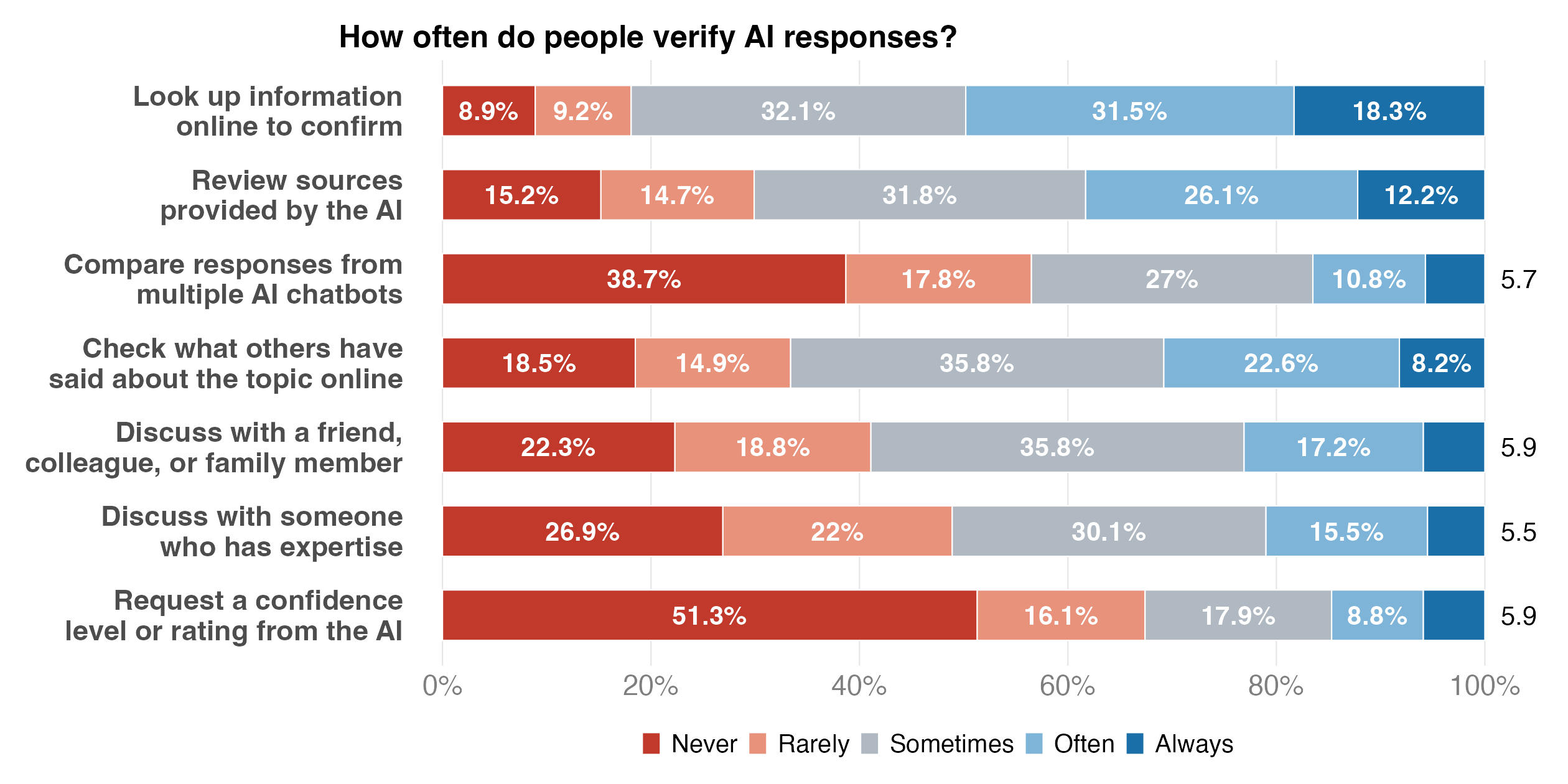}
    \caption{Prevalence of verification behaviours}
    \label{fig:survey:verification_behaviour}
\end{figure}

Overall, a large majority of respondents (94.7\%) indicated that they verified AI chatbot responses some (though not all) of the time, with 39 participants reporting that they never verified responses, and just 14 saying that they always verified responses.

The most common fact-checking behaviours involved looking up additional information online,  reviewing sources provided by the AI chatbot, and checking what other internet users have said, see Fig. \ref{fig:survey:verification_behaviour}.
The least common approaches were requesting a confidence level or rating from the AI chatbot itself and comparing responses from multiple AI chatbots.

\begin{figure*}[h]
    \centering
    \includegraphics[width=0.49\linewidth]{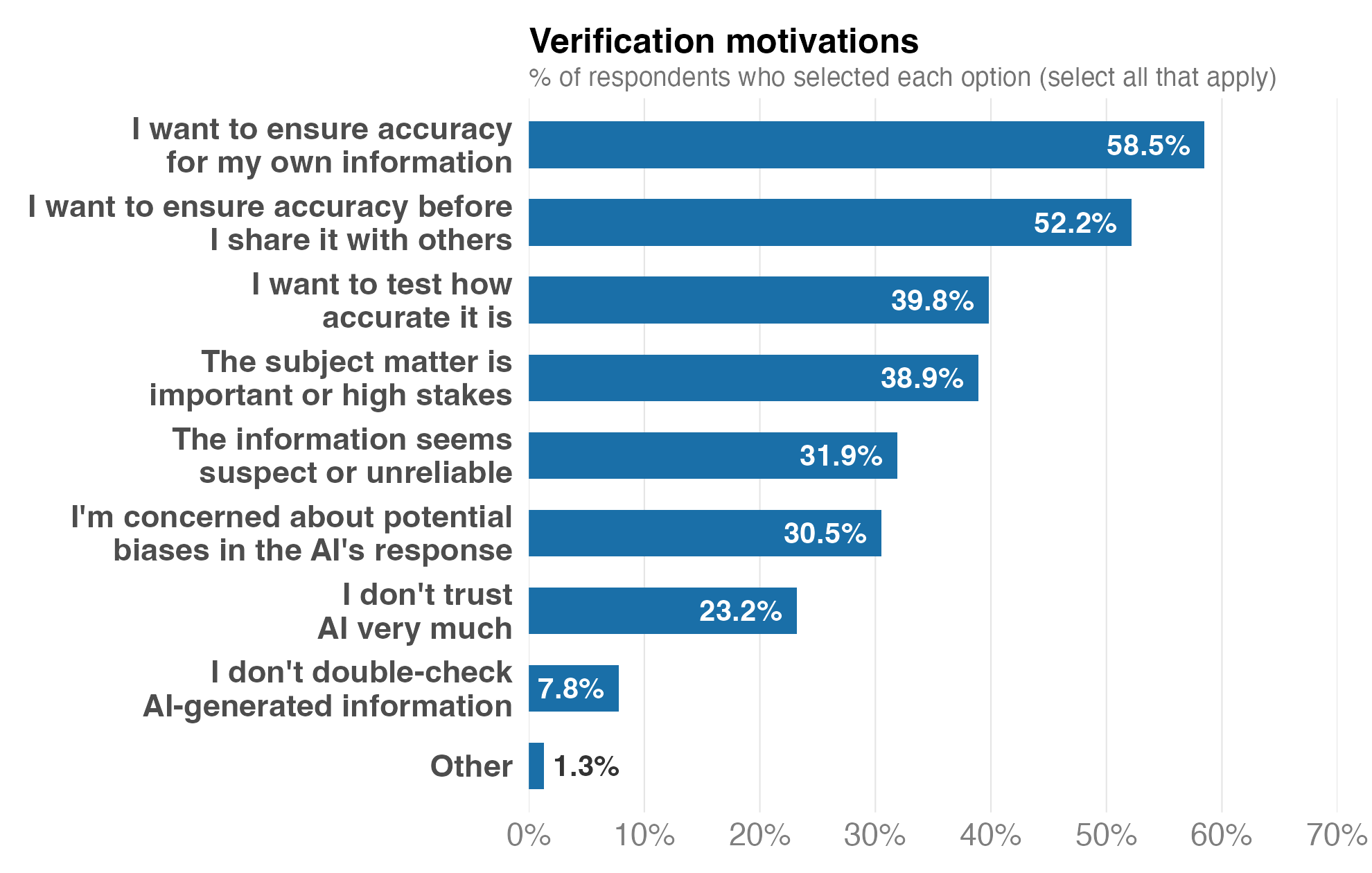}
    \includegraphics[width=0.49\linewidth]{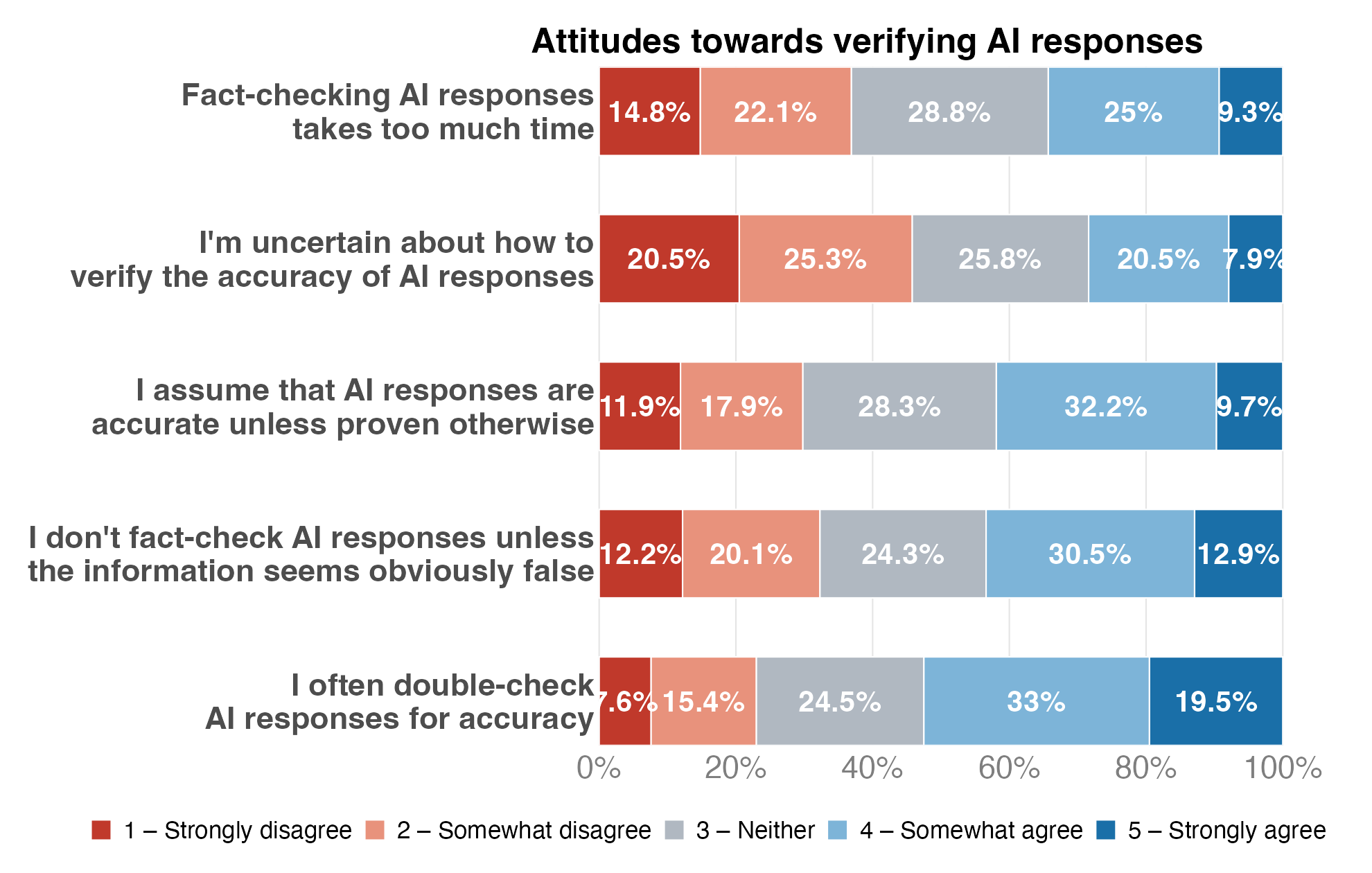}
    \caption{Motivations for verifying AI information (left) and Attitudes towards verification and fact-checking (right)}
    \label{fig:survey:verification_motivations}
\end{figure*}

When it came to motivations for verifying AI-generated information, the most common reasons involved ensuring accuracy, either for oneself (58.5\%), or before passing information on to others (52.2\%), or simply testing the accuracy of the AI-generated information (39.8\%) see Fig. \ref{fig:survey:verification_motivations}.
A substantial number (38.9\%) of people reported that high stakes subjects also prompted them to double-check AI responses.

People's attitudes towards verifying and fact-checking AI-generated responses were somewhat mixed and ambivalent (see Fig. \ref{fig:survey:verification_motivations}).
People reported that they often double-checked AI responses for accuracy to a moderate degree ($M = 3.41, SD = 1.18$), yet people neither agreed nor disagreed with statements such as ``I don't fact-check  AI responses unless the information seems obviously false'' ($M = 3.12, SD = 1.22$) and ``I assume that AI responses are accurate unless proven otherwise'' ($M = 3.10, SD = 1.16$).
People's attitudes towards the time it takes to fact-check AI responses ($M = 2.92, SD = 1.20$) and their uncertainty about how to verify the accuracy of AI responses ($M = 2.70, SD = 1.23$) also hovered around the middle of the scale.
These results suggest that although people see verification as important, their motivation and time and effort they are willing to expend on it may vary, perhaps depending on context and other factors.

Overall, the survey findings indicated that people placed importance on being able to evaluate the factuality of AI-generated information, especially for certain high-stakes queries, such as those involving medical or legal advice. However, people's verification behaviours and attitudes also indicated that people did not consistently or systematically verify information, and were only moderately satisfied with their own ability to evaluate AI responses, and with the information provided by current AI chatbots to facilitate this. In the next section, we conducted a large scale experiment to examine how factuality indicators shape people's perceptions of AI chatbots for information-seeking across various contexts.

\section{RQ3: How do factuality indicators affect perceptions of LLM chatbot factuality?}

Finally, we conducted a large-scale experiment in order to systematically compare factuality indicators and their impact on people's perceived accuracy, trust, and overall impressions of an AI chatbot across three different topics with varying stakes.

\subsection{Method}

\subsubsection{Participants}
The sample comprised 5,151 United States-based adults recruited through YouGov's online research panel
aged 18-89 ($M_{age}$ = 46.7, $SD_{age}$ = 15.8), who reported using generative AI, on a monthly basis or more frequently.
The sample was stratified based on gender, age, race and education, based on the 2023 Current Population Survey (CPS) to be representative of US adults.
All participants provided informed consent prior to the study.

\subsubsection{Design}
\label{RQ3:design}
We designed a between-participants experiment in which participants were tasked to evaluate an example of a conversation with a mobile AI chatbot (see Section \ref{RQ3:materials} for details).
Participants were randomly assigned to conditions in which the AI system they saw varied along three main factors: type of factuality indicator (\textit{No indicator}, \textit{Confidence-label}, \textit{Confidence-percentage}, \textit{Confidence bubble}, \textit{Source-label}, \textit{Source-label + confidence language}), topic (gardening, math, medical), and response accuracy (correct or incorrect).
People assigned to a condition involving a confidence indicator (\textit{Confidence-label}, \textit{Confidence-percentage}, \textit{Confidence-bubble}, \textit{Source label + confidence cue}) saw either a high confidence or a low confidence indicator.
Each participant saw only one type of example.

We measured participants' perceptions of the AI chatbot, including perceived accuracy and information sufficiency, as well as  trust and overall sentiment before and after the correct response was revealed to them, see Table \ref{tab:dependent_variables} for details.

\begin{table*}[t]
\centering \small
\begin{tabular}{p{1.85cm} p{1.8cm} p{6.3cm} p{4.0cm}} \toprule \textbf{Measure} & \textbf{Phase} & \textbf{Question} & \textbf{Scale} \\
\midrule Perceived Accuracy & Pre-revelation & “How accurate or inaccurate do you think the response from this AI chatbot is?” & 1 = Completely inaccurate, 5 = Completely accurate \\
Sufficiency of Information & Pre-revelation & “To what extent does this AI chatbot provide sufficient information to help you evaluate the accuracy of the response?” & 1 = Not at all, 5 = Very much \\
Trust & Pre- \& Post-revelation & “How much would you trust the responses from this AI chatbot?” & 1 = Not at all, 5 = Very much \\
Sentiment & Pre- \& Post-revelation & “Overall, how would you describe your impression of this AI chatbot?” & 1 = Very negative, 5 = Very positive \\
Intent to Use & Post-revelation & “How likely is it that you would use this AI chatbot?” & 1 = Not at all likely, 5 = Extremely likely
\\ Recommendability & Post-revelation & “How likely is it that you would recommend this AI chatbot to others?” & 1 = Not at all likely, 5 = Extremely likely \\ \bottomrule
\end{tabular}
\caption{Overview of dependent variables and measures in experiment}
\label{tab:dependent_variables}
\end{table*}

\subsubsection{Materials}
\label{RQ3:materials}
\begin{figure*}
    \centering
    \includegraphics[width=\linewidth]{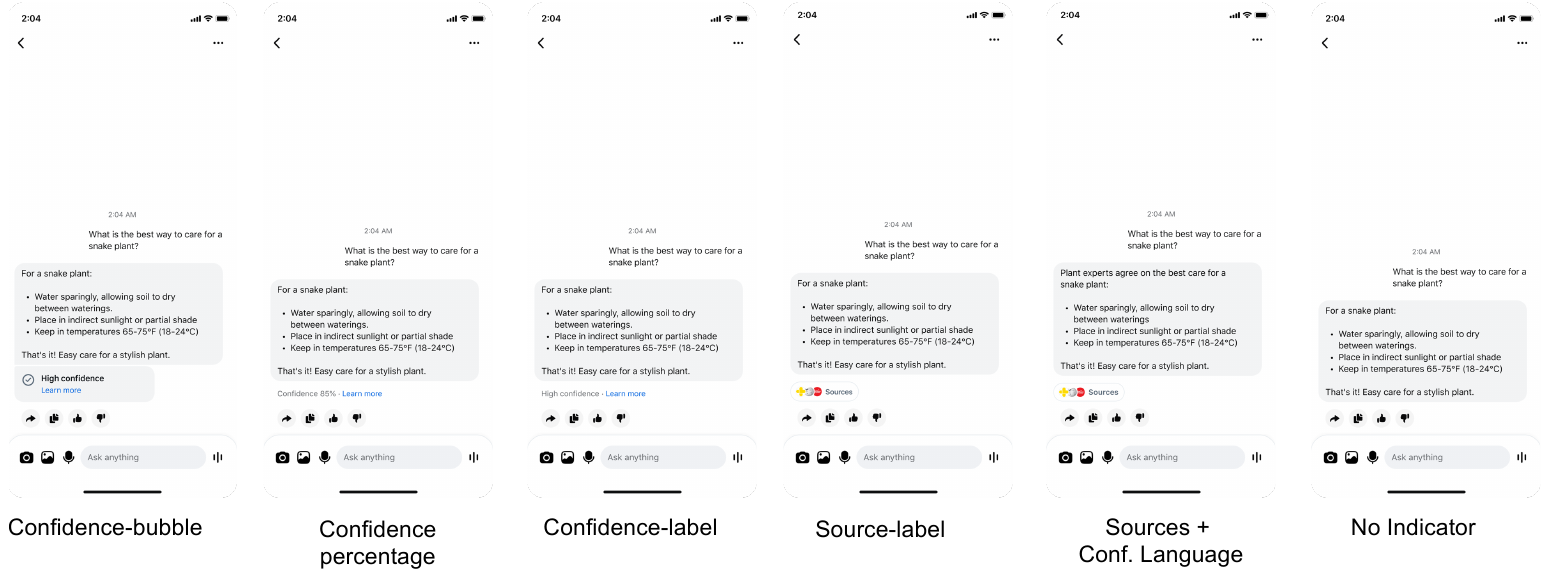}
    \caption{Examples of materials in Gardening conditions}
    \label{fig:RQ3materials}
\end{figure*}

We created a set of mobile chatbot conversation scenarios that varied in the factuality indicators cues presented to people and the objective correctness of the AI response.
All scenarios had the same structure: a user query followed by a chatbot-generated response.
We examined three different domains, which we chose based on the results of our survey and interviews to represent contexts in which people place high (medical advice), medium (education or work; a mathematical calculation), and lower (everyday task; gardening advice) importance on understanding how accurate an AI chatbot is.
The query and response belonged to one of three domains (gardening, math, or medical), and the response consisted of either a correct answer or an incorrect answer (see Table \ref{tab:expt_materials} for details).
We designed five different combinations of factuality indicators based on prior work and insights from RQ1 and RQ2, alongside a control condition (see Fig. \ref{fig:RQ3materials} for examples).

The \textit{Confidence-label} condition displayed a high- or low-confidence label directly beneath the chatbot response together with clickable text that read ``Learn more''.
The \textit{Confidence-bubble} condition presented the same confidence information (high/low) in a more visually prominent chat bubble beneath the response.
The \textit{Confidence-percentage} condition showed numerical confidence estimates (85\% for high confidence, 30\% for low confidence) beneath the response, instead of confidence labels.
In the \textit{Source-label} condition, participants were shown source icons beneath the model response.
In the \textit{Sources \& Confidence Language} condition, source icons were accompanied by a short textual statement suggesting either high confidence (e.g., ``Plant experts agree on this approach...'') or low confidence (e.g., ``Not many sources list these same instructions, so double-check with a trusted gardening resource'').
Finally, the \textit{No indicator} control condition showed just the chatbot response without any factuality indicators.

\begin{table*}[]
\begin{tabular}{lll}
\hline
\textbf{Gardening} &
  \textbf{Math} &
  \textbf{Medical} \\ \hline
\textbf{Query} &
   &
   \\ \hline
\begin{tabular}[c]{@{}l@{}}What is the best way to care for a \\ snake plant?\end{tabular} &
  \begin{tabular}[c]{@{}l@{}}A 2kg block is pushed with a force of 4N on a \\ frictionless surface. What is the acceleration \\ of the block?\end{tabular} &
  \begin{tabular}[c]{@{}l@{}}What is the recommended oral dosage of \\ Labetalol for treating high blood pressure?\end{tabular} \\ \hline
\textbf{Correct Response} &
   &
   \\ \hline
\begin{tabular}[c]{@{}l@{}}For a snake plant:\\ \\ Water sparingly, allowing soil to dry \\ between waterings\\ Place in indirect sunlight or partial shade\\ Keep in temperatures 65-75F (18-24C)\\ \\ That's it! Easy care for a stylish plant.\end{tabular} &
  \begin{tabular}[c]{@{}l@{}}To find the acceleration, we can use Newton's \\ second law of motion, which states that force (F) \\ equals mass (m) times acceleration (a):\\ F = ma\\ Given:\\ F = 4 N (force applied)\\ m = 2 kg (mass of the block)\\ We need to solve for acceleration (a).\\ Rearrange the formula:\\ a = F / m\\ a = 4 N / 2 kg\\ a = 2 m/s²\\ So, the acceleration of the block is 2 meters per \\ second squared\end{tabular} &
  \begin{tabular}[c]{@{}l@{}}The oral dosage recommendation for \\ Labetalol for hypertension typically starts \\ at 100 mg twice daily, with a usual \\ maintenance dose of 200-400 mg twice daily. \\ Doses can be adjusted based on patient \\ response, with some patients requiring up \\ to 1200-2400 mg per day in divided doses. \\ Consult a healthcare professional for \\ personalized guidance.\end{tabular} \\ \hline
\textbf{Incorrect Response} &
   &
   \\ \hline
\begin{tabular}[c]{@{}l@{}}For a snake plant:\\ \\ Water every day\\ Place in direct sunlight\\ Keep in temperatures 65-75F (18-24C)\\ \\ That's it! Easy care for a stylish plant.\end{tabular} &
  \begin{tabular}[c]{@{}l@{}}To find the acceleration, we can use Newton's \\ second law of motion, which states that \\ acceleration (a) equals mass (m) times force (F):\\ F = 4 N (force applied)\\ m = 2 kg (mass of the block)\\ a = F × m\\ a = 4 N × 2 kg\\ a = 8\\ So, the acceleration of the block is 8 meters per \\ second squared\end{tabular} &
  \begin{tabular}[c]{@{}l@{}}The oral dosage recommendation for \\ Labetalol for hypertension typically starts \\ at 500 mg twice daily, with a usual \\ maintenance dose of 1000-2000 mg twice \\ daily. Doses can be adjusted based on patient \\ response, with some patients requiring up \\ to 5000-7400 mg per day in divided doses. \\ Consult a healthcare professional for \\ personalized guidance.\end{tabular} \\ \hline
\textbf{High Confidence Response} &
   &
   \\ \hline
\begin{tabular}[c]{@{}l@{}}Plant experts agree on the best care for \\ a snake plant:\\ {[}...{]}\end{tabular} &
  \begin{tabular}[c]{@{}l@{}}{[}...{]} So according to this accepted approach, the\\ acceleration of the block is {[}2/8{]} meters per \\ second squared\end{tabular} &
  \begin{tabular}[c]{@{}l@{}}The widely accepted oral dosage \\ recommendation for Labetalol for \\ hypertension typically starts at {[}...{]}\end{tabular} \\ \hline
\textbf{Low Confidence Response} &
   &
   \\ \hline
\begin{tabular}[c]{@{}l@{}}{[}...{]}\\ That's it! Not many sources list these same \\ instructions, though, so double-check with \\ your favourite gardening resource.\end{tabular} &
  \begin{tabular}[c]{@{}l@{}}So the acceleration of the block is {[}2/8{]} per \\ second squared. There may be disagreement \\ in the math world, though, so double-check \\ this approach to finding acceleration.\end{tabular} &
  \begin{tabular}[c]{@{}l@{}}However, not all the sources I checked agree \\ on this information. Consult a healthcare \\ professional for personalized guidance.\end{tabular} \\ \hline
\end{tabular}
\caption{Topics and example stimulus content used in the experimental treatments.} \label{tab:expt_materials}
\end{table*}

\newcommand{\bluecheck}{{\color{blue}\checkmark}}

\subsubsection{Procedure}

Participants were provided with a link to the experiment, and first entered their demographic details.
Participants were shown an example of a conversation with a generative AI chatbot in the form of a smartphone screenshot, consisting of a user query and the AI chatbot's response, which varied depending on the experimental condition (see Fig. \ref{fig:RQ3materials}). They were required to review the conversation for at least 30 seconds before proceeding.
Then, they were asked a set of questions about their perceptions of the system (see Table \ref{tab:dependent_variables}).
Subsequently, on the next screen it was revealed whether the chatbot was correct or incorrect in its response, e.g., (''The correct response to the question `ABC' is `XYZ', meaning, the AI chatbot was [correct/incorrect].''
Following this revelation, participants were asked ``Knowing that the response was [correct/incorrect], please rate your final impression of the AI chatbot.'' Here, participants rated their overall impression of the chatbot and their trust in its responses.
Data collection took place in June 2025. The average completion time was 6m 27s.

\subsection{Results}

We analysed how factuality indicators influenced participants' perceptions of AI chatbot responses for different topics before and after ground-truth correctness was revealed.
Overall, we found that
(i) \textit{Confidence-labels} and \textit{Source-labels} scored highest on pre-revelation measures of accuracy, sufficiency, trust, and sentiment, though differences between indicators were relatively small, and
(ii) effects of indicators largely disappeared post-revelation, while response correctness had the largest effect on people's perceptions, and
(iii) the topic of responses played a significant role: participants' perceptions of responses for non-medical queries were more positive than for medical ones.

\begin{figure}[h]
     \centering
     \includegraphics[width=\linewidth]{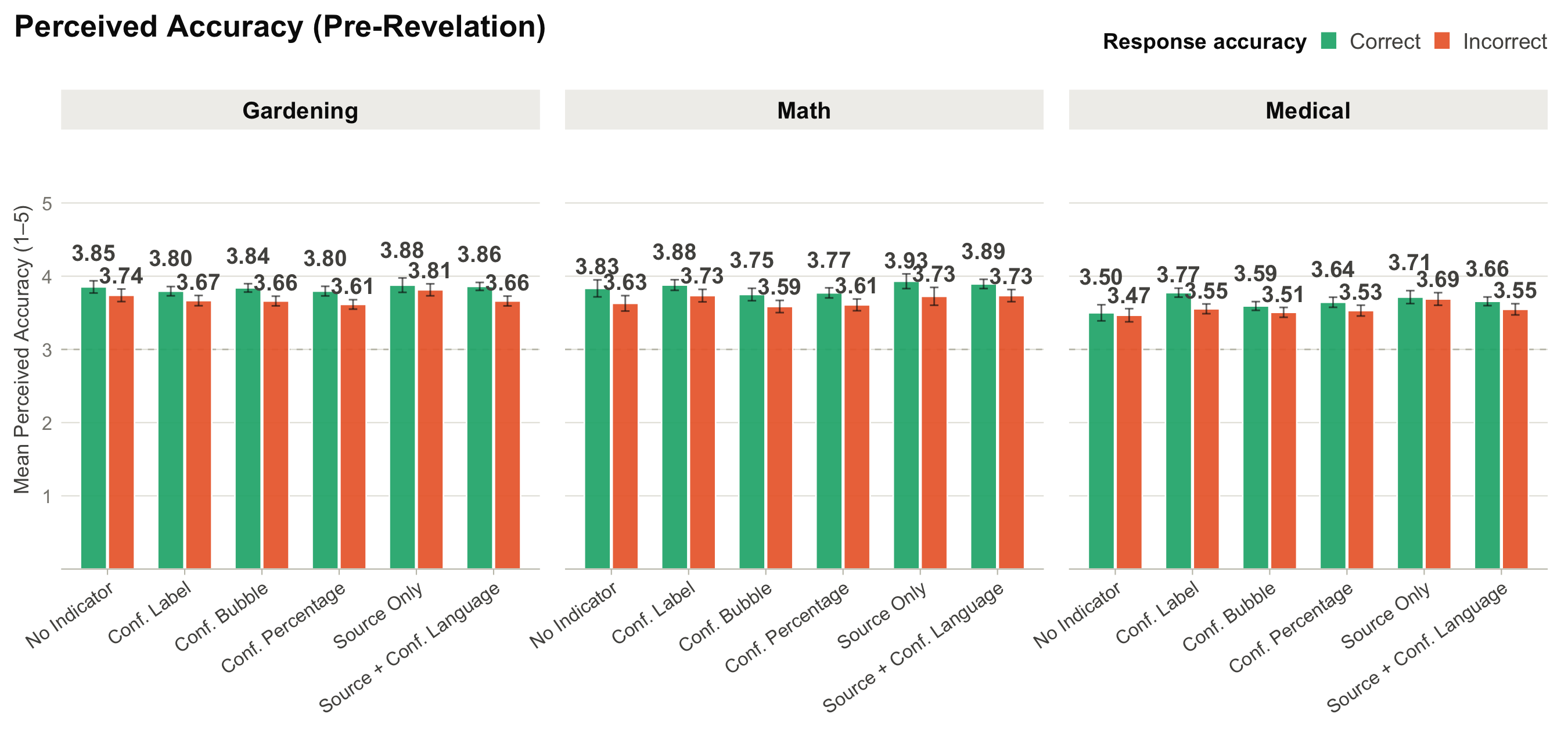}
     \caption{Participants' perceived accuracy judgments. Error bars represent standard error of the mean.}
     \label{fig:RQ3:perceived_acc}
 \end{figure}

\paragraph{Perceived accuracy.}
Participants' judgments of response accuracy were mainly influenced by both response correctness, F(1, 5115)=33.969, p <.001, $\eta_p^2$=.007 and topic F(2, 5115)=16.545, p <.001, $\eta_p^2$=.006, see Fig. \ref{fig:RQ3:perceived_acc}.
Correct responses (M=3.776) were perceived as significantly more accurate than incorrect responses (M=3.628) (p < .001, d = .161).
Topic also influenced perceived accuracy: gardening and mathematical responses were rated similarly (p = .996), while both gardening and mathematical responses received higher accuracy ratings than medical responses (both ps < .001).
There was also a main effect of indicator F(5, 5115)=2.527, p =.027, $\eta_p^2$=.002, however no significant differences emerged between the conditions after correcting for multiple pairwise comparisons.
\textit{Source-only} indicators (M=3.794) showed a trend toward higher perceived accuracy than both \textit{Confidence-bubble} indicators (M=3.658) (p = .067) and \textit{Confidence-percentage} indicators (M=3.661) (p = .072).

\begin{figure}[h]
     \centering
     \includegraphics[width=\linewidth]{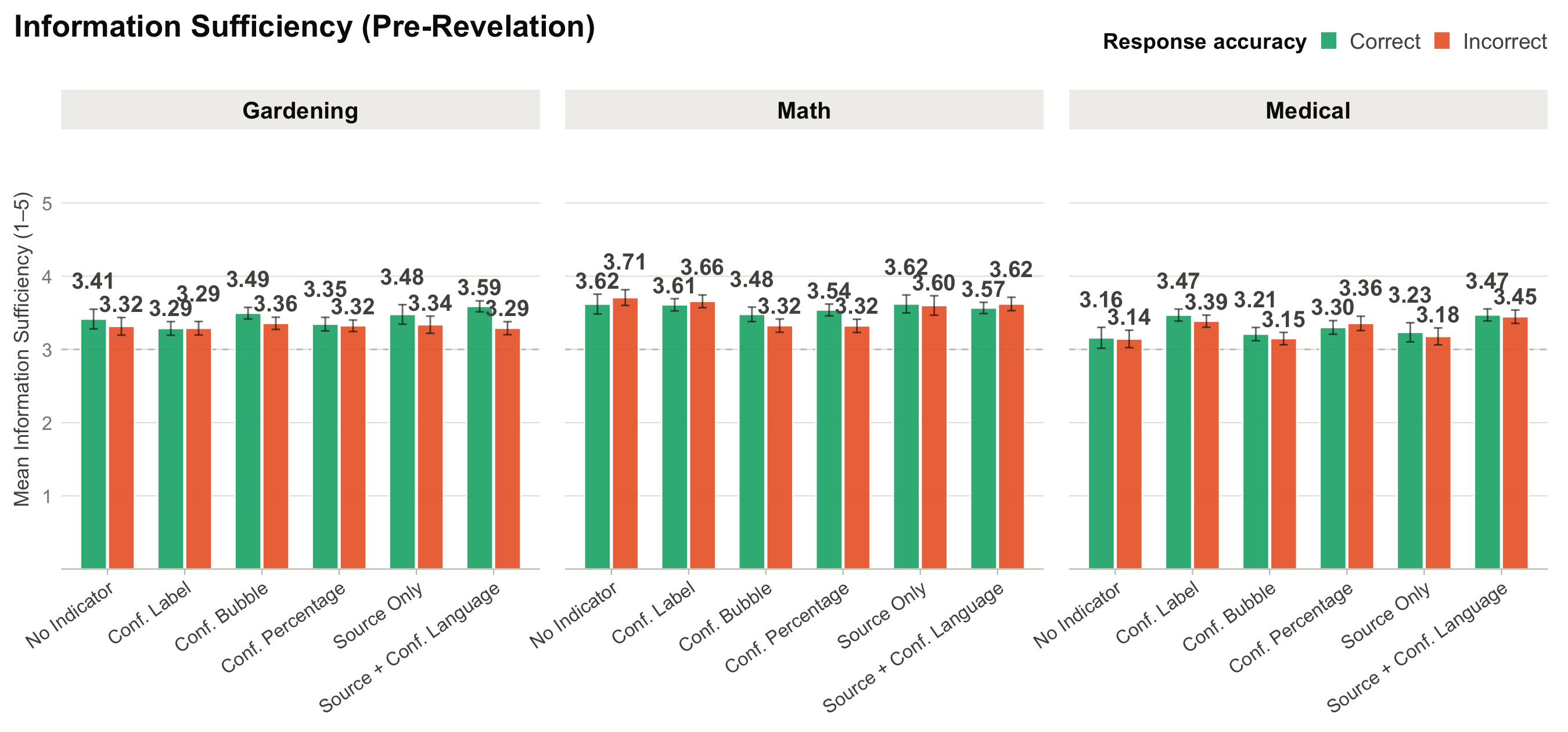}
     \caption{Participants' information sufficiency judgments. Error bars represent standard error of the mean.}
     \label{fig:RQ3:info_sufficiency}
 \end{figure}

\paragraph{Information sufficiency.}
There were main effects of indicator type, F(5, 5115)=2.565, p =.025, $\eta_p^2$=.002, correctness, F(1, 5115)=4.74, p =.0295, $\eta_p^2$=.001, and topic, F(2, 5115)=17.634, p <.001, $\eta_p^2$=.007 on people's perceptions of information sufficiency, see Fig. \ref{fig:RQ3:info_sufficiency}.
Correct responses (M=3.444) were associated with slightly higher sufficiency ratings than incorrect responses (M=3.374) (p = .039).
Topic also impacted perceived sufficiency. Mathematical responses (M=3.542) were judged as providing more sufficient information than both gardening (M=3.373) (p < .001) and medical (M=3.312) responses (p < .001), whereas no differences emerged between gardening and medical responses (p = .249).
People shown the \textit{Sources and Confidence-language} indicator (M=3.495) gave it the highest judgments of information sufficiency, significantly higher than those in the \textit{Confidence-bubble} condition (M=3.335), p = .012, d = .14.

\begin{figure}[h]
     \centering
     \includegraphics[width=\linewidth]{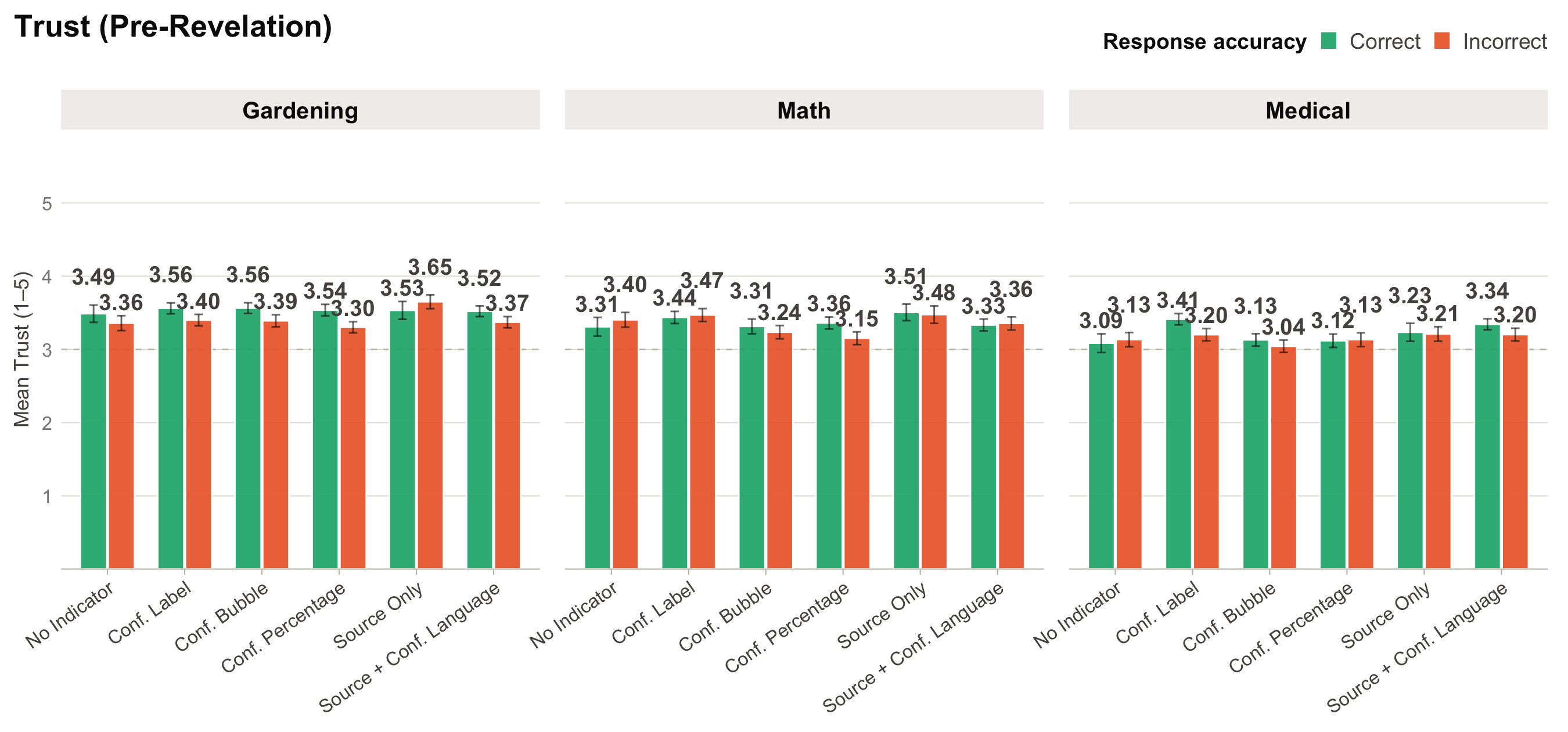}
     \includegraphics[width=\linewidth]{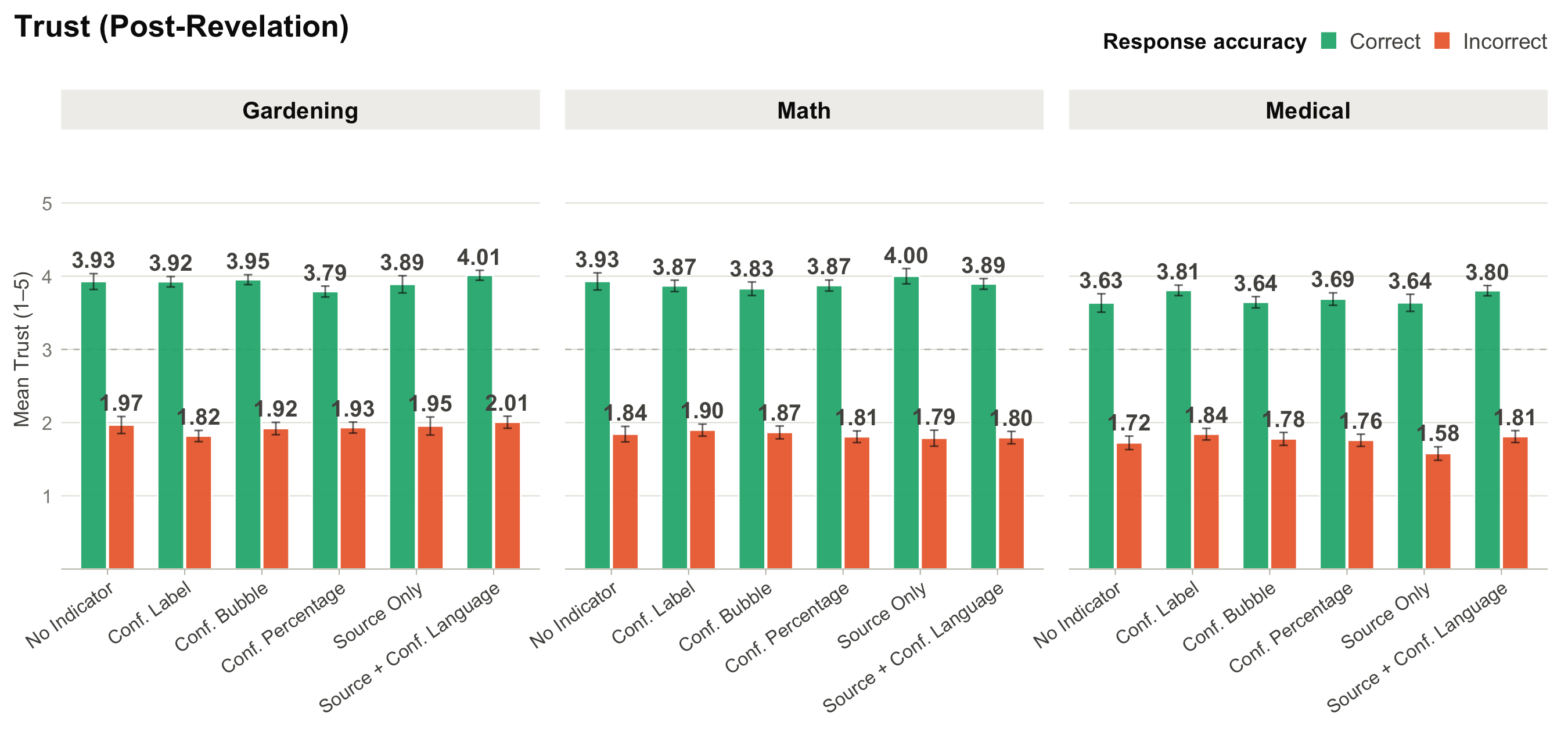}
     \caption{Participants' trust judgments, pre-accuracy revelation (above) and post-accuracy revelation (below). Error bars represent standard error of the mean.}
     \label{fig:RQ3:trust}
 \end{figure}

\paragraph{Pre-revelation: Perceived trust.}
Prior to the correctness of the AI system's response being revealed, there were main effects of indicator type, F(5, 5115)=3.38, p =.0048, $\eta_p^2$=.003, correctness, F(1, 5115)=9.177, p =.0025, $\eta_p^2$=.002, and topic, F(2, 5115)=28.087, p <.001, $\eta_p^2$=.011 on people's perceived trust judgments (see Fig. \ref{fig:RQ3:trust}).
\textit{Confidence-label} indicators received the highest initial trust ratings, while \textit{Source-label} indicators were the second highest, both were significantly higher than the lowest rated, \textit{confidence-percentage} indicators (p = .033, d = .132; p = .045, d = .157). No other differences between the indicators were significant.
Correct responses elicited greater trust than incorrect responses (p = .003, d = .084), although this effect was relatively small.
Initial trust also varied depending on topic: trust judgments for gardening responses were higher than both mathematical (p = .006, d = .105) and medical responses (p < .001, d = .254), and mathematical responses were trusted more than medical responses (p < .001, d = .149).

\paragraph{Post-revelation: Perceived trust.}
Once response correctness was revealed, the effect of factuality indicators on perceived trust disappeared F(5, 5115)=1.773, p =.115, see Fig. \ref{fig:RQ3:trust}.
The main effect of response correctness, F(1, 5115)=4836.01, p <.001, $\eta_p^2$=.485 was large: responses revealed as correct were trusted substantially more than responses revealed as incorrect (p < .001, d = 1.935).
The effect of topic also remained, F(2, 5115)=13.9, p<.001, $\eta_p^2$=.005.
Responses in the gardening and mathematical domains did not differ (p=.864), but were both still judged more trustworthy than in the medical domain (both ps<.001).

\begin{figure}[h]
     \centering
     \includegraphics[width=\linewidth]{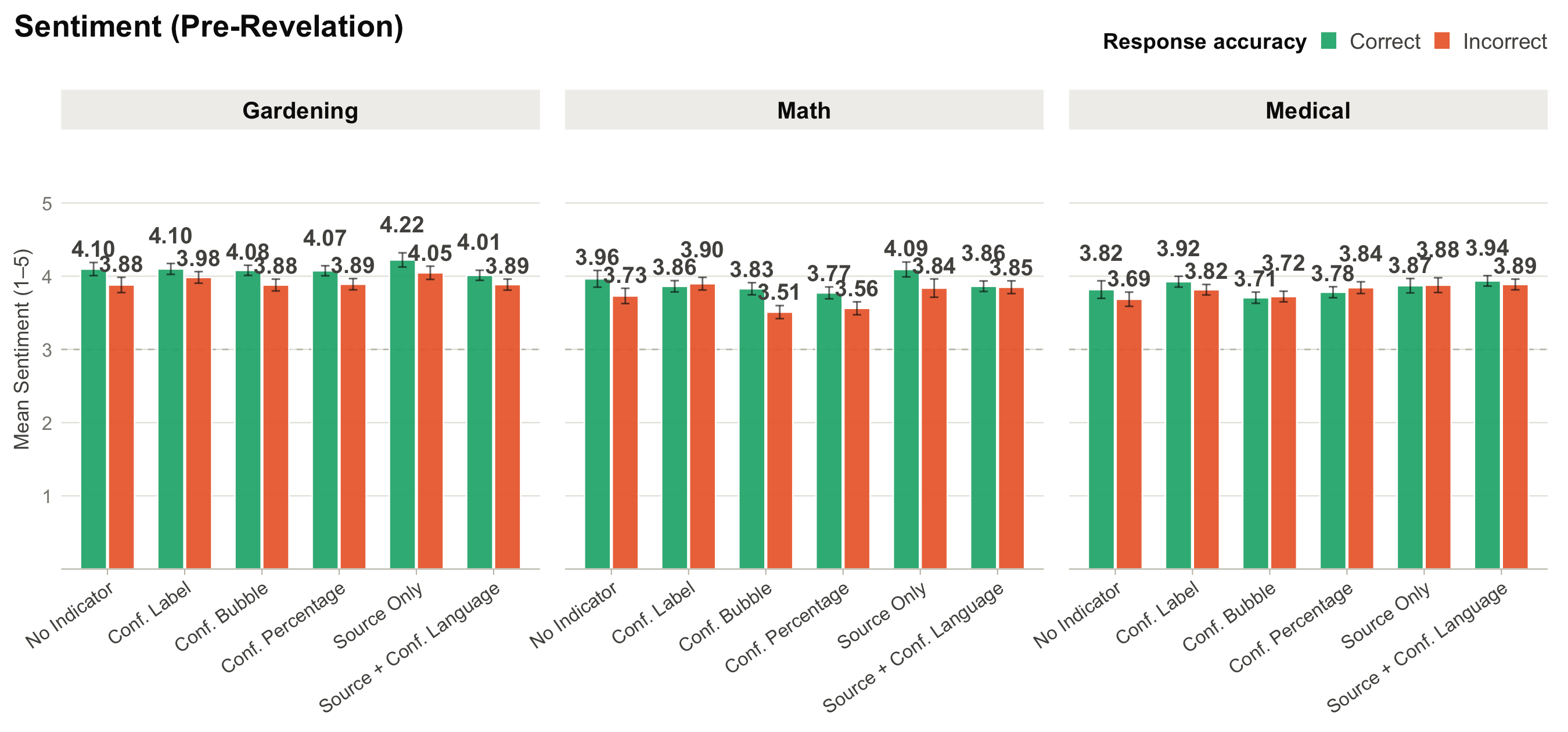}
     \includegraphics[width=\linewidth]{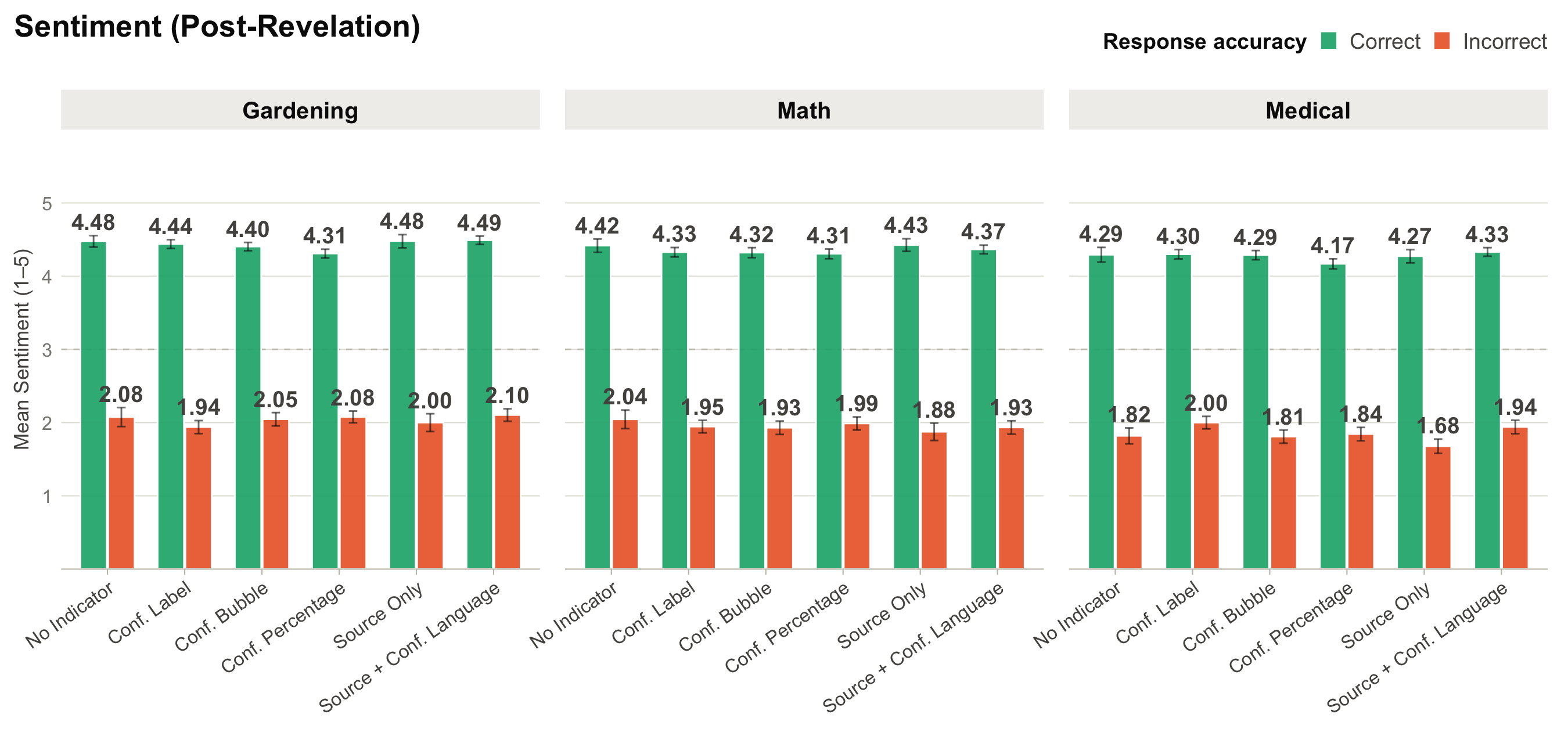}
     \caption{Participants' sentiment judgments, pre-accuracy revelation (above) and post-accuracy revelation (below). Error bars represent standard error of the mean.}
     \label{fig:RQ3:sentiment}
 \end{figure}

\paragraph{Pre-revelation: Sentiment.}
There were main effects of indicator type, F(5, 5115)=4.192, p <.001, $\eta_p^2$=.004, correctness, F(1, 5115)=16.449, p <.001, $\eta_p^2$=.003, and topic, F(2, 5115)=21.924, p <.001, $\eta_p^2$=.008 on people's overall impressions of the chatbot, see Fig. \ref{fig:RQ3:sentiment}.
\textit{Confidence-label} indicators generated more positive impressions than \textit{Confidence-bubble} indicators (p=.0318, d = .133), while \textit{Source-only} indicators generated more positive impressions than both \textit{Confidence-bubble} (p=.003, d = .2) and \textit{Confidence-percentage} (p=.015, d = .176) indicators.
Correct responses also elicited more positive impressions than incorrect responses (p<.001, d = .115).
Gardening responses received more positive evaluations than both mathematical and medical responses (p<.001, d=.21; p<.001, d=.179), while medical and mathematical responses did not differ (p=.729).

\paragraph{Post-revelation: Sentiment.}
Similar to trust, the effect of indicator type disappeared after the correctness of the model response was revealed, F(5, 5115)=1.697, p =.132, while correctness, F(1, 5115) =7412.75, p <.001, $\eta_p^2$=.591, and topic, F(2, 5115)=11.458, p <.001, $\eta_p^2$=.004 still influenced people's impressions of the chatbot, see Fig. \ref{fig:RQ3:sentiment}.
Aligning with the trust judgments, people unsurprisingly had more positive sentiment towards correct responses than incorrect ones (p<.001, d = 2.395), and gardening and mathematical responses were evaluated more positively than medical responses (p<.001, d = .162; p<.001, d = .13), while gardening and mathematical responses did not differ from one another (p=.628).

\subsubsection{Effects of confidence levels}
We also examined the impact of high and low confidence level indicators on  participants' judgments during the experiment.
For these analyses, only participants who were shown some sort of confidence indicator (i.e., \textit{Confidence-bubble}, \textit{Confidence percentage}, \textit{Confidence-label}, and \textit{Sources \& Confidence language}) were included.

\paragraph{Perceived accuracy}
We conducted a 2 (accuracy: correct vs incorrect) x 2 (confidence: high vs low) x 3 (topic: gardening, math, medical) x 4 (confidence indicator: Confidence-bubble, Confidence percentage, Confidence-label, and Sources \& Confidence language) ANOVA on people's judgments of perceived accuracy.

There was a main effect of confidence level, F(1, 4059) = 29.396, p <.001, $\eta^2_p$ = .007.
People perceived responses labelled with high confidence ($M = 3.77, SD = .896$) to be more accurate than those labelled as low confidence ($M = 3.62, SD = .922$)
There was also a main effect of correctness, F(1, 4059) = 30.393, p <.001, $\eta^2_p$ = .007.
People perceived correct responses ($M = 3.773, SD = .848$) to be more accurate than incorrect responses ($M = 3.618, SD = .965$).
There was a main effect of topic, F(2, 4059) = 10.973, p <.001, $\eta^2_p$ = .005.
People perceived Medical ($M = 3.6, SD = .87$) responses to be less accurate than both Gardening ($M = 3.734, SD = .851$), p <.001, d = .16 and Math responses ($M = 3.749, SD = 1.005$), p <.001, d = .16.
There was no difference in how accurate people perceived Gardening and Math responses, p = .91.
There was no effect of factuality indicator, F(3, 4059) = 2.118, p = .096.
There were no significant interactions between the factors, p > .05 for all comparisons.

\paragraph{Information sufficiency}
We conducted a 2 (accuracy: correct vs incorrect) x 2 (confidence: high vs low) x 3 (topic: gardening, math, medical) x 4 (confidence indicator: Confidence-bubble, Confidence-percentage, Confidence-label, and Sources \& Confidence-language) ANOVA on people's judgments of information sufficiency.
There was no effect of confidence level, F(1, 4059) = 1.267, p = .261 (High M = 3.43, SD = 1.141, Low M = 3.392, SD = 1.143).

There was a main effect of correctness, F(1, 4059) = 4.223, p = .04, $\eta^2_p$ = .001.
People judged that correct responses ($M = 3.448, SD = 1.119$) contained more sufficient information to evaluate response accuracy than incorrect responses ($M = 3.375, SD = 1.163$).

There was a main effect of topic, F(2, 4059) = 8.61, p <.001, $\eta^2_p$ = .004.
People gave higher judgments of information sufficiency to
Math responses ($M = 3.517, SD = 1.135$) than Medical ($M = 3.348, SD = 1.148$), p <.001, d = .15 and Gardening responses ($M = 3.37, SD = 1.141$), p = .002, d = .13.
There was a main effect of factuality indicator, F(3, 4059) = 4.053, p = .007, $\eta^2_p$ = .003.
There was a significant interaction between the topic and indicator factors, F(6, 4059) = 2.423, p = .0244, $\eta^2_p$ = .004.
No other interactions were significant, p > .05 for all comparisons.

\subsubsection{Effects of confidence indicator misalignment on subjective trust and sentiment change}
We also investigated the impact of the alignment between confidence indicator (i.e., high or low) and the correctness of the AI response on people's trust and sentiment judgments pre- and post-revelation. We conducted a 2 (accuracy: correct vs incorrect) x 2 (confidence: high vs low) x 3 (topic: gardening, math, medical) x 4 (confidence indicator: Confidence-bubble, Confidence-percentage, Confidence-label, and Sources \& Confidence-language) ANOVA.

\paragraph{Trust}
There was a main effect of confidence level F(1, 4059) = 14.519, p <.001, $\eta^2_p$ = .004.
Participants' trust judgments decreased more when they had been shown high confidence responses ($M = -.566, SD = 1.416$), than when they had been shown low confidence responses ($M = -.412, SD = 1.363$), p <.001, d = .11.
There was also a main effect of correctness F(1, 4059) = 3437.497, p <.001, $\eta^2_p$ = .456.
Participants' trust increased when responses were revealed to be correct ($M = .454, SD = .772$), and decreased when they were revealed to be incorrect ($M = -1.419, SD = 1.231$).
There was a main effect of topic F(2, 4059) = 3.57, p = .0282, $\eta^2_p$ = .002. However, posthoc Tukey HSD tests showed the trust changes for each topic (Gardening: $M = -.563, SD = 1.35$; Math: $M = -.462, SD = 1.46$; Medical: $M = -.44, SD = 1.361$) were not significantly different after correcting for multiple comparisons.
There was no effect of indicator, F(3, 4059) = 2.359, p = .07.
There was a significant interaction between confidence level and correctness, F(1, 4059) = 5.057, p = .0246, $\eta^2_p$ = .0012.
There were no other significant interactions p >.05 for all comparisons.

\paragraph{Sentiment}
There was a main effect of confidence F(1, 4059) = 18.429, p <.001, $\eta^2_p$ = .005.
Participants' overall sentiment decreased more for high confidence responses ($M = -.813, SD = 1.638$), than they did for low confidence responses ($M = -.625, SD = 1.554$).
There was also a main effect of correctness F(1, 4059) = 4265.804, p <.001, $\eta^2_p$ = .508.
Participants' overall impression of the system improved when responses were revealed to be correct ($M = .427, SD = .821$), and decreased when they were revealed to be incorrect ($M = -1.848, SD = 1.35$).
There was a main effect of topic F(2, 4059) = 5.819, p = .003, $\eta^2_p$ = .003. Posthoc Tukey HSD tests showed that sentiment dropped less in the Math topic ($M = -.609, SD = 1.621$) than it did for Gardening ($M = -.795, SD = 1.567$), p = .006, d = .117. Math did not differ from Medical ($M = -.749, SD = 1.605$), p = .06, nor did Gardening differ from Math, p =.73.
There was no effect of indicator, F(3, 4059) = 1.94, p = .121.
There was a significant interaction between confidence level and correctness, F(1, 4059) = 6.307, p = .0121, $\eta^2_p$ = .0015.
There was also a significant interaction between topic and indicator, F(6, 4059) = 2.43, p = .024, $\eta^2_p$ = .0036.
Topic did not interact with confidence or correctness, and the other factors did not interact, p >.05 for all comparisons.

\begin{figure*}
    \centering
    \includegraphics[width=\linewidth]{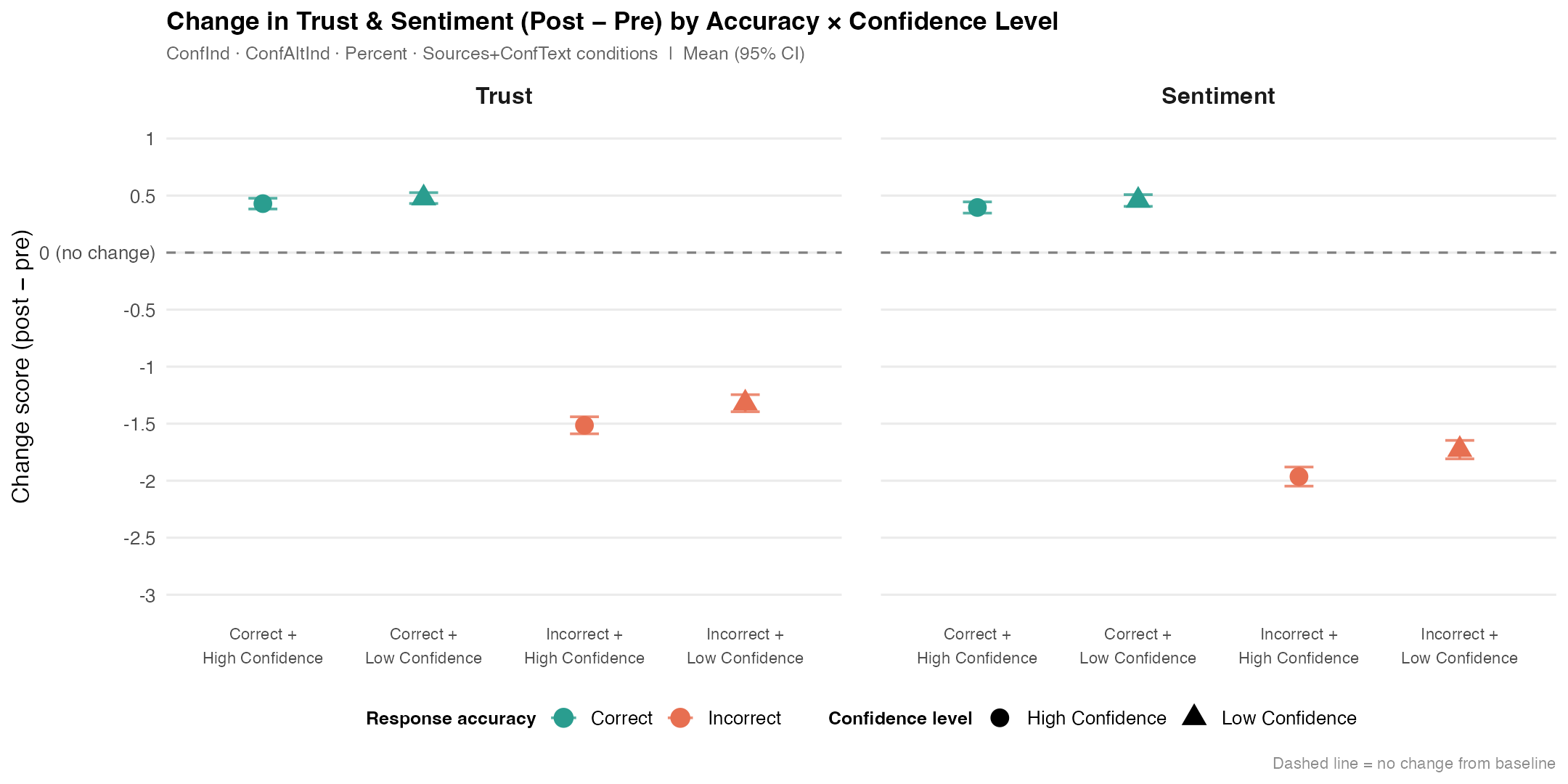}
    \caption{Change in trust and sentiment (post-pre) by response correctness and confidence level}
    \label{fig:placeholder}
\end{figure*}

\section{Discussion}
Our mixed-methods investigation reveals that while users express a strong desire for factuality indicators to help them assess AI-generated responses, the effectiveness of these cues is heavily moderated by AI response correctness, task context, interface design, and subjective interpretations of confidence.

\subsection{The role of source and confidence indicators in assessing factuality}
Although our survey and interview participants expressed that factuality indicators for AI chatbots would be useful, our experiment indicated that the impact of these interventions on people's perceptions in the experiment was relatively small, and affected only initial trust and sentiment towards the system response.
We found that people who were provided with source indicators gave higher judgments of information sufficiency and initial overall sentiment than people shown the \textit{Confidence-bubble} indicator. \textit{Confidence-label} and \textit{Source-label} indicators led to higher pre-revelation trust.
However, we also observed that any significant differences between indicators in the pre-accuracy revelation measures disappeared once participants were informed about whether the AI response was correct or not.
Perceived accuracy, trust, and sentiment also often did not differ from the control condition in which no factuality indicators were provided. Moreover, the sizes of the effects attributed to factuality indicators were small, indicating that these interventions may have limited impact in practice. This may have been due to the between-participants design of the experiment, in which participants only saw one type of indicator, rather than directly comparing them, which might have elicited stronger preferences.
Together, these results highlight that while factuality indicators may influence people's initial impressions, response accuracy and being able to access reliable information play the greatest role in shaping people's perceptions of AI systems.

\subsection{The importance of aligning and communicating confidence accurately}
Our interview participants consistently stated that they wanted AI systems to communicate confidence. In the experiment, confidence indicators influenced pre-revelation judgments of trust, sentiment, and perceived accuracy (although the effect sizes were small), with high-confidence signals generally leading to more favourable evaluations than low-confidence signals.
However, in the experiment, we also found that misaligned (high) confidence indicators for incorrect responses led to more extreme drops in trust than for misaligned (low) confidence for correct responses.
This pattern echoes work on algorithm aversion, which shows that people lose trust in an algorithm quickly once they see it make mistakes \cite{dietvorst2015algorithm,dejong2025algorithm}. A confidently presented error may be judged particularly harshly because it breaks the expectation the indicator itself created.
This asymmetry also suggests that conservative confidence indicators may be preferable when using chatbots for information-seeking, as people perceive the consequences of following incorrect advice as more severe than not following correct advice.
This finding also underscores the importance of accurate confidence calibration for model developers: including factuality indicators that are not reliable undermines trust.

Moreover, our interviews surfaced that participants held diverse and sometimes contradictory interpretations of what confidence actually meant. Some interpreted confidence as the probability of correctness, others as agreement among sources, and some as evidence that the AI had already verified its answer. Several participants shifted between interpretations during the same interview.
This finding highlights a significant transparency issue, as well as a risk for overreliance: each interpretation implies a different guarantee, e.g., a user who reads `85\% confident' as `already verified' is likely to under-verify relative to a user who reads it as `85\% likely correct.'
Our findings also align with broader research on uncertainty communication, which has repeatedly shown that numerical and verbal uncertainty expressions are interpreted in multiple ways by different people \cite{wallsten1993preferences,dhami2005interpersonal}, and echoes recent calls for closer alignment between human notions of uncertainty and how it is typically operationalised in AI models \cite{ulmer2026anthropomimetic,warren2025explainablefactchecking}.
One solution to this could be to provide clearer explanations to users of what given confidence scores represent and how to interpret them e.g., likelihood, or faithfulness to the underlying source.
Another could be to provide explanations of how confidence scores are derived or calculated, by explicitly connecting confidence estimates to underlying evidence, agreement among sources, retrieval quality, or other contributing factors. For example, several participants described wanting an explanation of what confidence scores were based on, and one participant explicitly proposed highlighting the specific span of a response that was less certain rather than (or in addition to) a single confidence score for the entire response.
Prior work has shown that highlighting can help people to validate reliability, and have positive impacts on trust \cite{do2025highlight,Vasconcelos2025generationprob}.
Such approaches could build on methods of explaining the uncertainty associated with an output by highlighting the parts of the evidence that contribute to it (e.g., \cite{sun-etal-2026-explaining}).

\subsection{Sources are necessary, but not sufficient alone}
Across both the survey and interview studies, participants repeatedly emphasised the importance of sources. People reported reviewing sources more often than requesting confidence scores, and interview participants frequently described sources as evidence that could be independently inspected and validated.
In our experiment, participants who were shown source indicators along with confidence-aligned language gave the highest information sufficiency judgments.
However, our findings also suggest that merely providing sources alone is insufficient to support factuality assessment.
Some participants admitted that even when sources are offered, clicking through to inspect them can feel like too much effort, mirroring previous findings that, although people believe that sources should be available, they can be reluctant to engage with them \cite{kim2024LLMuncertainty,kim2025fostering,vasconcelos_explanations_2023}.
Our findings also point to potential ways of designing such systems to facilitate more verification.
Presenting source information on demand through integrated, dropdown interface designs instead of pop-ups, which can appear disconnected from the AI response, could reduce the friction often involved in verifying information \cite{warren2026evidence}.
Interview participants also suggested that presenting direct relevant excerpts or pull-quotes from the source material could reduce the cognitive effort required to verify information and helped users easily identify potential hallucinations.
Further work in human-AI interaction, natural language processing, and information retrieval is therefore needed to develop methods of allowing users to verify AI sources, while managing the effort needed to evaluate them.

\subsection{Considering stakes when designing factuality indicators}
Finally, a recurring theme across our three studies was the role of context in how AI responses are perceived and trusted.
Survey respondents reported that knowing AI uncertainty was most important for health, legal, and financial topics, and interview participants repeatedly highlighted high-stakes domains as where factuality indicators were most critical.
In the experiment, differences between domains also emerged consistently: medical responses elicited lower trust, lower sentiment, and lower perceived accuracy than mathematics or everyday gardening questions.
Furthermore, confidence indicators also appeared less effective for medical topics,  with weaker effects than on other topics.
On one hand, it seems somewhat counterintuitive that factuality indicators appear to have less impact for contexts in which people emphasise that they are particularly important. One possible explanation for this may be that people are often guided by intuition in high stakes scenarios involving potential risk  \cite{slovic2010affect}.
This finding indicates that people weigh factuality indicators alongside perceived risk or stakes, and that confidence indicators that may be sufficient for everyday questions may be inadequate for queries involving health, or finances.
Future work is required to examine in more detail how factuality indicators should be designed and presented in context-sensitive ways.
For example, high-stakes contexts may warrant richer evidence displays, and greater transparency about uncertainty, while low-stakes contexts may benefit from more minimal, less disruptive indicators.

\subsection{Limitations and future work}
Our findings are subject to several limitations, and point to some promising areas of future work.
First, all three studies recruited only US-based participants, and required participants to already use generative AI tools regularly, and so our findings may not generalise to other populations.
Although the survey (RQ2) and experiment (RQ3) targeted a representative sample of US adults, the participants in our interview sample were all relatively young (under 30) and therefore our insights for this part of the study were limited to the experiences of this younger demographic.
Future work is therefore needed to examine how people in different socio-geographical contexts and those less familiar with generative AI assess AI factuality and perceived factuality indicators and what expectations they may have.
Our participants were also drawn from  opt-in online panels, which may be less representative of the population than address-based sampling or other random methods.

Second, our experiment only showed participants static screenshots of an AI chatbot interaction, rather than facilitating more realistic interactions with a chatbot. This limited the ecological validity of our design in several ways, for example, potentially reducing the effort and/or friction of actually clicking a source or a confidence explanation in comparison to a real life scenario.
We also focused on participants' perceptions of the system, rather than behavioural measures (e.g., reliance \cite{kim2024LLMuncertainty,kim2025fostering,schemmer2023relianceexplanations}) or decisions (e.g., deciding to share or publish an AI-generated response \cite{warren2026evidence}), as other studies have done. This may have contributed to the small effect sizes we observed in the experiment; participants may have had less reason to engage with the factuality indicators and responses, as no follow up action or decision was required.
Another limitation of our design was that each participant saw only one example of a conversation between chatbot and user, and so their judgments are based on this single impression rather than repeated exposure over multiple examples as they might in everyday life. Nevertheless, our results shed light on how initial impressions are formed and shaped by AI accuracy.
Examining how various factuality indicators affect how people interact and make decisions with AI chatbots in real-life settings, and how perceptions and reliance are shaped over time is a critical area for future work.

\section{Conclusion}
We examined how people assess AI factuality and how different factuality indicators influence their perceptions through a mixed-methods investigation comprising qualitative interviews, an observational survey, and a large-scale experiment.
Our findings shed light on how people integrate confidence cues, sources, and contextual considerations when deciding whether information is reliable.
Our interview findings highlight how providing factuality indicators such as confidence and sources can support this process, but they must be clearly interpretable and easily accessible to users, as well as being sensitive to contextual factors like topic and complexity.
Furthermore, our experimental results show that people ultimately judge systems based on their accuracy, and suggest that rather than maximising trust or sentiment, effective factuality indicators should help people to form well-calibrated expectations and make informed decisions about when and how to verify AI-generated information.

\begin{acks}
\noindent $\begin{array}{l}\includegraphics[width=1cm]{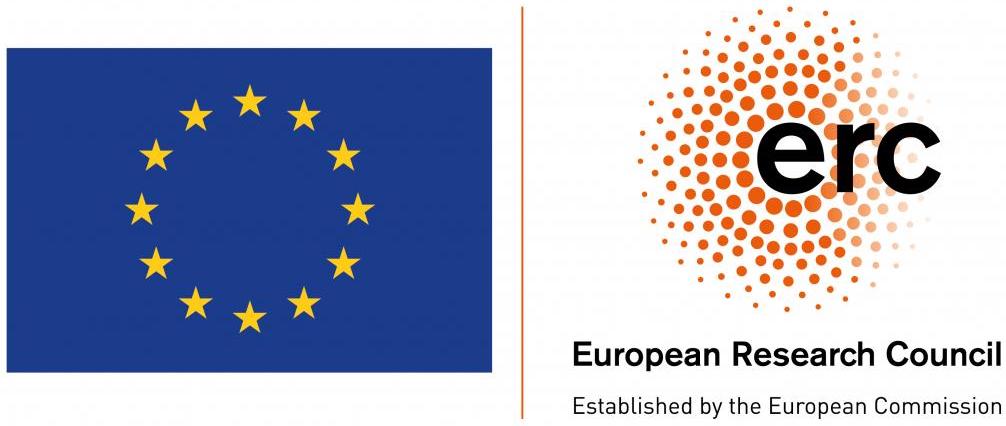} \end{array}$ This research was co-funded by the European Union (ERC, ExplainYourself, 101077481), and supported by the Pioneer Centre for AI, DNRF grant number P1. Views and opinions expressed are those of the authors only and do not necessarily reflect those of the European Union or the European Research Council. Neither the European Union nor the granting authority can be held responsible for them.

\end{acks}

\bibliographystyle{ACM-Reference-Format}
\bibliography{2_references}

%%% -*-BibTeX-*-
%%% Do NOT edit. File created by BibTeX with style
%%% ACM-Reference-Format-Journals [18-Jan-2012].

\begin{thebibliography}{41}

%%% ====================================================================
%%% NOTE TO THE USER: you can override these defaults by providing
%%% customized versions of any of these macros before the \bibliography
%%% command.  Each of them MUST provide its own final punctuation,
%%% except for \shownote{} and \showURL{}.  The latter two
%%% do not use final punctuation, in order to avoid confusing it with
%%% the Web address.
%%%
%%% To suppress output of a particular field, define its macro to expand
%%% to an empty string, or better, \unskip, like this:
%%%
%%% \newcommand{\showURL}[1]{\unskip}   % LaTeX syntax
%%%
%%% \def \showURL #1{\unskip}           % plain TeX syntax
%%%
%%% ====================================================================

\ifx \showCODEN    \undefined \def \showCODEN     #1{\unskip}     \fi
\ifx \showISBNx    \undefined \def \showISBNx     #1{\unskip}     \fi
\ifx \showISBNxiii \undefined \def \showISBNxiii  #1{\unskip}     \fi
\ifx \showISSN     \undefined \def \showISSN      #1{\unskip}     \fi
\ifx \showLCCN     \undefined \def \showLCCN      #1{\unskip}     \fi
\ifx \shownote     \undefined \def \shownote      #1{#1}          \fi
\ifx \showarticletitle \undefined \def \showarticletitle #1{#1}   \fi
\ifx \showURL      \undefined \def \showURL       {\relax}        \fi
% The following commands are used for tagged output and should be
% invisible to TeX
\providecommand\bibfield[2]{#2}
\providecommand\bibinfo[2]{#2}
\providecommand\natexlab[1]{#1}
\providecommand\showeprint[2][]{arXiv:#2}

\bibitem[Alkaissi and McFarlane(2023)]%
        {alkaissi2023artificial}
\bibfield{author}{\bibinfo{person}{Hussam Alkaissi} {and} \bibinfo{person}{Samy~I McFarlane}.} \bibinfo{year}{2023}\natexlab{}.
\newblock \showarticletitle{Artificial hallucinations in ChatGPT: implications in scientific writing}.
\newblock \bibinfo{journal}{\emph{Cureus}} \bibinfo{volume}{15}, \bibinfo{number}{2} (\bibinfo{year}{2023}).
\newblock
\href{https://doi.org/10.7759/cureus.35179}{doi:\nolinkurl{10.7759/cureus.35179}}


\bibitem[Augenstein et~al\mbox{.}(2024)]%
        {augenstein2024factualityLLMs}
\bibfield{author}{\bibinfo{person}{Isabelle Augenstein}, \bibinfo{person}{Timothy Baldwin}, \bibinfo{person}{Meeyoung Cha}, \bibinfo{person}{Tanmoy Chakraborty}, \bibinfo{person}{Giovanni~Luca Ciampaglia}, \bibinfo{person}{David Corney}, \bibinfo{person}{Renee DiResta}, \bibinfo{person}{Emilio Ferrara}, \bibinfo{person}{Scott Hale}, \bibinfo{person}{Alon Halevy}, \bibinfo{person}{Eduard Hovy}, \bibinfo{person}{Heng Ji}, \bibinfo{person}{Filippo Menczer}, \bibinfo{person}{Ruben Miguez}, \bibinfo{person}{Preslav Nakov}, \bibinfo{person}{Dietram Scheufele}, \bibinfo{person}{Shivam Sharma}, {and} \bibinfo{person}{Giovanni Zagni}.} \bibinfo{year}{2024}\natexlab{}.
\newblock \showarticletitle{Factuality challenges in the era of large language models and opportunities for fact-checking}.
\newblock \bibinfo{journal}{\emph{Nature Machine Intelligence}} (\bibinfo{year}{2024}), \bibinfo{pages}{1--12}.
\newblock
\href{https://doi.org/10.1038/s42256-024-00881-z}{doi:\nolinkurl{10.1038/s42256-024-00881-z}}


\bibitem[Bansal et~al\mbox{.}(2021)]%
        {bansal2021teamperformance}
\bibfield{author}{\bibinfo{person}{Gagan Bansal}, \bibinfo{person}{Tongshuang Wu}, \bibinfo{person}{Joyce Zhou}, \bibinfo{person}{Raymond Fok}, \bibinfo{person}{Besmira Nushi}, \bibinfo{person}{Ece Kamar}, \bibinfo{person}{Marco~Tulio Ribeiro}, {and} \bibinfo{person}{Daniel Weld}.} \bibinfo{year}{2021}\natexlab{}.
\newblock \showarticletitle{Does the Whole Exceed its Parts? The Effect of AI Explanations on Complementary Team Performance}. In \bibinfo{booktitle}{\emph{Proceedings of the 2021 CHI Conference on Human Factors in Computing Systems}} (Yokohama, Japan) \emph{(\bibinfo{series}{CHI '21})}. \bibinfo{publisher}{Association for Computing Machinery}, \bibinfo{address}{New York, NY, USA}, Article \bibinfo{articleno}{81}, \bibinfo{numpages}{16}~pages.
\newblock
\showISBNx{9781450380966}
\href{https://doi.org/10.1145/3411764.3445717}{doi:\nolinkurl{10.1145/3411764.3445717}}


\bibitem[Bel{\'e}m et~al\mbox{.}(2024)]%
        {belem-etal-2024-perceptions}
\bibfield{author}{\bibinfo{person}{Catarina~G. Bel{\'e}m}, \bibinfo{person}{Markelle Kelly}, \bibinfo{person}{Mark Steyvers}, \bibinfo{person}{Sameer Singh}, {and} \bibinfo{person}{Padhraic Smyth}.} \bibinfo{year}{2024}\natexlab{}.
\newblock \showarticletitle{Perceptions of Linguistic Uncertainty by Language Models and Humans}. In \bibinfo{booktitle}{\emph{Proceedings of the 2024 Conference on Empirical Methods in Natural Language Processing}}, \bibfield{editor}{\bibinfo{person}{Yaser Al-Onaizan}, \bibinfo{person}{Mohit Bansal}, {and} \bibinfo{person}{Yun-Nung Chen}} (Eds.). \bibinfo{publisher}{Association for Computational Linguistics}, \bibinfo{address}{Miami, Florida, USA}, \bibinfo{pages}{8467--8502}.
\newblock
\href{https://doi.org/10.18653/v1/2024.emnlp-main.483}{doi:\nolinkurl{10.18653/v1/2024.emnlp-main.483}}


\bibitem[Bhatt et~al\mbox{.}(2021)]%
        {bhatt2021uncertaintytransparency}
\bibfield{author}{\bibinfo{person}{Umang Bhatt}, \bibinfo{person}{Javier Antor\'{a}n}, \bibinfo{person}{Yunfeng Zhang}, \bibinfo{person}{Q.~Vera Liao}, \bibinfo{person}{Prasanna Sattigeri}, \bibinfo{person}{Riccardo Fogliato}, \bibinfo{person}{Gabrielle Melan\c{c}on}, \bibinfo{person}{Ranganath Krishnan}, \bibinfo{person}{Jason Stanley}, \bibinfo{person}{Omesh Tickoo}, \bibinfo{person}{Lama Nachman}, \bibinfo{person}{Rumi Chunara}, \bibinfo{person}{Madhulika Srikumar}, \bibinfo{person}{Adrian Weller}, {and} \bibinfo{person}{Alice Xiang}.} \bibinfo{year}{2021}\natexlab{}.
\newblock \showarticletitle{Uncertainty as a Form of Transparency: Measuring, Communicating, and Using Uncertainty}. In \bibinfo{booktitle}{\emph{Proceedings of the 2021 AAAI/ACM Conference on AI, Ethics, and Society}} (Virtual Event, USA) \emph{(\bibinfo{series}{AIES '21})}. \bibinfo{publisher}{Association for Computing Machinery}, \bibinfo{address}{New York, NY, USA}, \bibinfo{pages}{401–413}.
\newblock
\showISBNx{9781450384735}
\href{https://doi.org/10.1145/3461702.3462571}{doi:\nolinkurl{10.1145/3461702.3462571}}


\bibitem[Bo et~al\mbox{.}(2025)]%
        {bo2025relianceinterventions}
\bibfield{author}{\bibinfo{person}{Jessica~Y Bo}, \bibinfo{person}{Sophia Wan}, {and} \bibinfo{person}{Ashton Anderson}.} \bibinfo{year}{2025}\natexlab{}.
\newblock \showarticletitle{To Rely or Not to Rely? Evaluating Interventions for Appropriate Reliance on Large Language Models}. In \bibinfo{booktitle}{\emph{Proceedings of the 2025 CHI Conference on Human Factors in Computing Systems}} \emph{(\bibinfo{series}{CHI '25})}. \bibinfo{publisher}{Association for Computing Machinery}, \bibinfo{address}{New York, NY, USA}, Article \bibinfo{articleno}{905}, \bibinfo{numpages}{23}~pages.
\newblock
\showISBNx{9798400713941}
\href{https://doi.org/10.1145/3706598.3714097}{doi:\nolinkurl{10.1145/3706598.3714097}}


\bibitem[Bu\c{c}inca et~al\mbox{.}(2025)]%
        {bucinca2025decisionskills}
\bibfield{author}{\bibinfo{person}{Zana Bu\c{c}inca}, \bibinfo{person}{Siddharth Swaroop}, \bibinfo{person}{Amanda~E. Paluch}, \bibinfo{person}{Finale Doshi-Velez}, {and} \bibinfo{person}{Krzysztof~Z. Gajos}.} \bibinfo{year}{2025}\natexlab{}.
\newblock \showarticletitle{Contrastive Explanations That Anticipate Human Misconceptions Can Improve Human Decision-Making Skills}. In \bibinfo{booktitle}{\emph{Proceedings of the 2025 CHI Conference on Human Factors in Computing Systems}} \emph{(\bibinfo{series}{CHI '25})}. \bibinfo{publisher}{Association for Computing Machinery}, \bibinfo{address}{New York, NY, USA}, Article \bibinfo{articleno}{1024}, \bibinfo{numpages}{25}~pages.
\newblock
\showISBNx{9798400713941}
\href{https://doi.org/10.1145/3706598.3713229}{doi:\nolinkurl{10.1145/3706598.3713229}}


\bibitem[Ceron et~al\mbox{.}(2026)]%
        {ceron2026wildseekevaluatinglanguagemodels}
\bibfield{author}{\bibinfo{person}{Tanise Ceron}, \bibinfo{person}{Joachim Baumann}, \bibinfo{person}{Elisa Bassignana}, \bibinfo{person}{Berat Cabuk}, \bibinfo{person}{Dirk Hovy}, {and} \bibinfo{person}{Debora Nozza}.} \bibinfo{year}{2026}\natexlab{}.
\newblock \bibinfo{title}{WildSEEK: Evaluating Language Models for Information-Seeking}.
\newblock
\showeprint[arxiv]{2608.30683}~[cs.CL]
\urldef\tempurl%
\url{https://arxiv.org/abs/2608.30683}
\showURL{%
\tempurl}


\bibitem[Chiaburu et~al\mbox{.}(2024)]%
        {chiaburu2024uncertainty}
\bibfield{author}{\bibinfo{person}{Teodor Chiaburu}, \bibinfo{person}{Frank Hau{\ss}er}, {and} \bibinfo{person}{Felix Bie{\ss}mann}.} \bibinfo{year}{2024}\natexlab{}.
\newblock \showarticletitle{Uncertainty in xai: Human perception and modeling approaches}.
\newblock \bibinfo{journal}{\emph{Machine Learning and Knowledge Extraction}} \bibinfo{volume}{6}, \bibinfo{number}{2} (\bibinfo{year}{2024}), \bibinfo{pages}{1170--1192}.
\newblock
\href{https://doi.org/10.3390/make6020055}{doi:\nolinkurl{10.3390/make6020055}}


\bibitem[de~Jong et~al\mbox{.}(2025)]%
        {dejong2025algorithm}
\bibfield{author}{\bibinfo{person}{Sander de Jong}, \bibinfo{person}{Maarten~W Bos}, \bibinfo{person}{Niels van Berkel}, {and} \bibinfo{person}{Maarten~H Lamers}.} \bibinfo{year}{2025}\natexlab{}.
\newblock \showarticletitle{Algorithm appreciation or aversion: the effects of accuracy disclosure on users' reliance on algorithmic suggestions}.
\newblock \bibinfo{journal}{\emph{Behaviour \& Information Technology}} (\bibinfo{year}{2025}), \bibinfo{pages}{1--20}.
\newblock
\href{https://doi.org/10.1080/0144929X.2025.2535732}{doi:\nolinkurl{10.1080/0144929X.2025.2535732}}


\bibitem[Dhami and Wallsten(2005)]%
        {dhami2005interpersonal}
\bibfield{author}{\bibinfo{person}{Mandeep~K Dhami} {and} \bibinfo{person}{Thomas~S Wallsten}.} \bibinfo{year}{2005}\natexlab{}.
\newblock \showarticletitle{Interpersonal comparison of subjective probabilities: Toward translating linguistic probabilities}.
\newblock \bibinfo{journal}{\emph{Memory \& cognition}}  \bibinfo{volume}{33} (\bibinfo{year}{2005}), \bibinfo{pages}{1057--1068}.
\newblock
\href{https://doi.org/10.3758/BF03193213}{doi:\nolinkurl{10.3758/BF03193213}}


\bibitem[Dietvorst et~al\mbox{.}(2015)]%
        {dietvorst2015algorithm}
\bibfield{author}{\bibinfo{person}{Berkeley~J Dietvorst}, \bibinfo{person}{Joseph~P Simmons}, {and} \bibinfo{person}{Cade Massey}.} \bibinfo{year}{2015}\natexlab{}.
\newblock \showarticletitle{Algorithm aversion: people erroneously avoid algorithms after seeing them err.}
\newblock \bibinfo{journal}{\emph{Journal of experimental psychology: General}} \bibinfo{volume}{144}, \bibinfo{number}{1} (\bibinfo{year}{2015}), \bibinfo{pages}{114}.
\newblock
\href{https://doi.org/10.1037/xge0000033}{doi:\nolinkurl{10.1037/xge0000033}}


\bibitem[Do and Geyer(2025)]%
        {DoGeyer2025hideorhighlight}
\bibfield{author}{\bibinfo{person}{Hyo~Jin Do} {and} \bibinfo{person}{Werner Geyer}.} \bibinfo{year}{2025}\natexlab{}.
\newblock \showarticletitle{Hide or Highlight: Understanding the Impact of Factuality Expression on User Trust}.
\newblock \bibinfo{journal}{\emph{Proceedings of the AAAI/ACM Conference on AI, Ethics, and Society}} \bibinfo{volume}{8}, \bibinfo{number}{1}, \bibinfo{pages}{785--797}.
\newblock
\href{https://doi.org/10.1609/aies.v8i1.36589}{doi:\nolinkurl{10.1609/aies.v8i1.36589}}


\bibitem[Do et~al\mbox{.}(2025)]%
        {do2025highlight}
\bibfield{author}{\bibinfo{person}{Hyo~Jin Do}, \bibinfo{person}{Rachel Ostrand}, \bibinfo{person}{Werner Geyer}, \bibinfo{person}{Keerthiram Murugesan}, \bibinfo{person}{Dennis Wei}, {and} \bibinfo{person}{Justin Weisz}.} \bibinfo{year}{2025}\natexlab{}.
\newblock \showarticletitle{Highlight All the Phrases: Enhancing LLM Transparency through Visual Factuality Indicators}. In \bibinfo{booktitle}{\emph{Proceedings of the AAAI/ACM Conference on AI, Ethics, and Society}}, Vol.~\bibinfo{volume}{8}. \bibinfo{pages}{772--784}.
\newblock
\href{https://doi.org/10.1609/aies.v8i1.36588}{doi:\nolinkurl{10.1609/aies.v8i1.36588}}


\bibitem[Grossman et~al\mbox{.}(2026)]%
        {grossman2026AIdisruptssearch}
\bibfield{author}{\bibinfo{person}{Riley Grossman}, \bibinfo{person}{Songjiang Liu}, \bibinfo{person}{Michael~K. Chen}, \bibinfo{person}{Mike Smith}, \bibinfo{person}{Cristian Borcea}, {and} \bibinfo{person}{Yi Chen}.} \bibinfo{year}{2026}\natexlab{}.
\newblock \showarticletitle{How Generative AI Disrupts Search: An Empirical Study of Google Search, Gemini, and AI Overviews}. In \bibinfo{booktitle}{\emph{Proceedings of the 49th International ACM SIGIR Conference on Research and Development in Information Retrieval}} (Melbourne, VIC, Australia) \emph{(\bibinfo{series}{SIGIR '26})}. \bibinfo{publisher}{Association for Computing Machinery}, \bibinfo{address}{New York, NY, USA}, \bibinfo{pages}{448–459}.
\newblock
\showISBNx{9798400725999}
\href{https://doi.org/10.1145/3805712.3809667}{doi:\nolinkurl{10.1145/3805712.3809667}}


\bibitem[Hofman et~al\mbox{.}(2020)]%
        {hofman2020uncertainty}
\bibfield{author}{\bibinfo{person}{Jake~M. Hofman}, \bibinfo{person}{Daniel~G. Goldstein}, {and} \bibinfo{person}{Jessica Hullman}.} \bibinfo{year}{2020}\natexlab{}.
\newblock \showarticletitle{How Visualizing Inferential Uncertainty Can Mislead Readers About Treatment Effects in Scientific Results}. In \bibinfo{booktitle}{\emph{Proceedings of the 2020 CHI Conference on Human Factors in Computing Systems}} (Honolulu, HI, USA) \emph{(\bibinfo{series}{CHI '20})}. \bibinfo{publisher}{Association for Computing Machinery}, \bibinfo{address}{New York, NY, USA}, \bibinfo{pages}{1–12}.
\newblock
\showISBNx{9781450367080}
\href{https://doi.org/10.1145/3313831.3376454}{doi:\nolinkurl{10.1145/3313831.3376454}}


\bibitem[Ibrahim et~al\mbox{.}(2025)]%
        {ibrahim2025measuringmitigatingoverreliancenecessary}
\bibfield{author}{\bibinfo{person}{Lujain Ibrahim}, \bibinfo{person}{Katherine~M. Collins}, \bibinfo{person}{Sunnie S.~Y. Kim}, \bibinfo{person}{Anka Reuel}, \bibinfo{person}{Max Lamparth}, \bibinfo{person}{Kevin Feng}, \bibinfo{person}{Lama Ahmad}, \bibinfo{person}{Prajna Soni}, \bibinfo{person}{Alia~El Kattan}, \bibinfo{person}{Merlin Stein}, \bibinfo{person}{Siddharth Swaroop}, \bibinfo{person}{Ilia Sucholutsky}, \bibinfo{person}{Andrew Strait}, \bibinfo{person}{Q.~Vera Liao}, {and} \bibinfo{person}{Umang Bhatt}.} \bibinfo{year}{2025}\natexlab{}.
\newblock \bibinfo{title}{Measuring and mitigating overreliance is necessary for building human-compatible AI}.
\newblock
\showeprint[arxiv]{2509.08010}~[cs.CY]
\urldef\tempurl%
\url{https://arxiv.org/abs/2509.08010}
\showURL{%
\tempurl}


\bibitem[Ji et~al\mbox{.}(2023)]%
        {ji2023hallucination}
\bibfield{author}{\bibinfo{person}{Ziwei Ji}, \bibinfo{person}{Nayeon Lee}, \bibinfo{person}{Rita Frieske}, \bibinfo{person}{Tiezheng Yu}, \bibinfo{person}{Dan Su}, \bibinfo{person}{Yan Xu}, \bibinfo{person}{Etsuko Ishii}, \bibinfo{person}{Ye~Jin Bang}, \bibinfo{person}{Andrea Madotto}, {and} \bibinfo{person}{Pascale Fung}.} \bibinfo{year}{2023}\natexlab{}.
\newblock \showarticletitle{Survey of Hallucination in Natural Language Generation}.
\newblock \bibinfo{journal}{\emph{ACM Comput. Surv.}} \bibinfo{volume}{55}, \bibinfo{number}{12}, Article \bibinfo{articleno}{248} (\bibinfo{date}{March} \bibinfo{year}{2023}), \bibinfo{numpages}{38}~pages.
\newblock
\showISSN{0360-0300}
\href{https://doi.org/10.1145/3571730}{doi:\nolinkurl{10.1145/3571730}}


\bibitem[Kim et~al\mbox{.}(2024)]%
        {kim2024LLMuncertainty}
\bibfield{author}{\bibinfo{person}{Sunnie S.~Y. Kim}, \bibinfo{person}{Q.~Vera Liao}, \bibinfo{person}{Mihaela Vorvoreanu}, \bibinfo{person}{Stephanie Ballard}, {and} \bibinfo{person}{Jennifer~Wortman Vaughan}.} \bibinfo{year}{2024}\natexlab{}.
\newblock \showarticletitle{"I'm Not Sure, But...": Examining the Impact of Large Language Models' Uncertainty Expression on User Reliance and Trust}. In \bibinfo{booktitle}{\emph{Proceedings of the 2024 ACM Conference on Fairness, Accountability, and Transparency}} (Rio de Janeiro, Brazil) \emph{(\bibinfo{series}{FAccT '24})}. \bibinfo{publisher}{Association for Computing Machinery}, \bibinfo{address}{New York, NY, USA}, \bibinfo{pages}{822–835}.
\newblock
\showISBNx{9798400704505}
\href{https://doi.org/10.1145/3630106.3658941}{doi:\nolinkurl{10.1145/3630106.3658941}}


\bibitem[Kim et~al\mbox{.}(2025)]%
        {kim2025fostering}
\bibfield{author}{\bibinfo{person}{Sunnie S.~Y. Kim}, \bibinfo{person}{Jennifer~Wortman Vaughan}, \bibinfo{person}{Q.~Vera Liao}, \bibinfo{person}{Tania Lombrozo}, {and} \bibinfo{person}{Olga Russakovsky}.} \bibinfo{year}{2025}\natexlab{}.
\newblock \showarticletitle{Fostering Appropriate Reliance on Large Language Models: The Role of Explanations, Sources, and Inconsistencies}. In \bibinfo{booktitle}{\emph{Proceedings of the 2025 CHI Conference on Human Factors in Computing Systems}} \emph{(\bibinfo{series}{CHI '25})}. \bibinfo{publisher}{Association for Computing Machinery}, \bibinfo{address}{New York, NY, USA}, Article \bibinfo{articleno}{420}, \bibinfo{numpages}{19}~pages.
\newblock
\showISBNx{9798400713941}
\href{https://doi.org/10.1145/3706598.3714020}{doi:\nolinkurl{10.1145/3706598.3714020}}


\bibitem[Kirsten et~al\mbox{.}(2026)]%
        {kirsten-etal-2026-characterizing}
\bibfield{author}{\bibinfo{person}{Elisabeth Kirsten}, \bibinfo{person}{Jost~Gro{\ss}e Perdekamp}, \bibinfo{person}{Qinyuan Wu}, \bibinfo{person}{Mihir Upadhyay}, \bibinfo{person}{Krishna~P. Gummadi}, {and} \bibinfo{person}{Muhammad~Bilal Zafar}.} \bibinfo{year}{2026}\natexlab{}.
\newblock \showarticletitle{Characterizing Web Search in The Age of Generative {AI}}. In \bibinfo{booktitle}{\emph{Findings of the {A}ssociation for {C}omputational {L}inguistics: {ACL} 2026}}, \bibfield{editor}{\bibinfo{person}{Maria Liakata}, \bibinfo{person}{Viviane~P. Moreira}, \bibinfo{person}{Jiajun Zhang}, {and} \bibinfo{person}{David Jurgens}} (Eds.). \bibinfo{publisher}{Association for Computational Linguistics}, \bibinfo{address}{San Diego, California, United States}, \bibinfo{pages}{10827--10848}.
\newblock
\showISBNx{979-8-89176-395-1}
\href{https://doi.org/10.18653/v1/2026.findings-acl.526}{doi:\nolinkurl{10.18653/v1/2026.findings-acl.526}}


\bibitem[Lee and See(2004)]%
        {lee2004trust}
\bibfield{author}{\bibinfo{person}{John~D Lee} {and} \bibinfo{person}{Katrina~A See}.} \bibinfo{year}{2004}\natexlab{}.
\newblock \showarticletitle{Trust in automation: Designing for appropriate reliance}.
\newblock \bibinfo{journal}{\emph{Human factors}} \bibinfo{volume}{46}, \bibinfo{number}{1} (\bibinfo{year}{2004}), \bibinfo{pages}{50--80}.
\newblock
\href{https://doi.org/10. 1518/hfes.46.1.50_30392}{doi:\nolinkurl{10. 1518/hfes.46.1.50_30392}}


\bibitem[Liu et~al\mbox{.}(2023)]%
        {liu-etal-2023-evaluating}
\bibfield{author}{\bibinfo{person}{Nelson Liu}, \bibinfo{person}{Tianyi Zhang}, {and} \bibinfo{person}{Percy Liang}.} \bibinfo{year}{2023}\natexlab{}.
\newblock \showarticletitle{Evaluating Verifiability in Generative Search Engines}. In \bibinfo{booktitle}{\emph{Findings of the Association for Computational Linguistics: EMNLP 2023}}, \bibfield{editor}{\bibinfo{person}{Houda Bouamor}, \bibinfo{person}{Juan Pino}, {and} \bibinfo{person}{Kalika Bali}} (Eds.). \bibinfo{publisher}{Association for Computational Linguistics}, \bibinfo{address}{Singapore}, \bibinfo{pages}{7001--7025}.
\newblock
\href{https://doi.org/10.18653/v1/2023.findings-emnlp.467}{doi:\nolinkurl{10.18653/v1/2023.findings-emnlp.467}}


\bibitem[Miller(2023)]%
        {miller2023explainable}
\bibfield{author}{\bibinfo{person}{Tim Miller}.} \bibinfo{year}{2023}\natexlab{}.
\newblock \showarticletitle{Explainable AI is Dead, Long Live Explainable AI! Hypothesis-driven Decision Support using Evaluative AI}. In \bibinfo{booktitle}{\emph{Proceedings of the 2023 ACM Conference on Fairness, Accountability, and Transparency}} (Chicago, IL, USA) \emph{(\bibinfo{series}{FAccT '23})}. \bibinfo{publisher}{Association for Computing Machinery}, \bibinfo{address}{New York, NY, USA}, \bibinfo{pages}{333–342}.
\newblock
\showISBNx{9798400701924}
\href{https://doi.org/10.1145/3593013.3594001}{doi:\nolinkurl{10.1145/3593013.3594001}}


\bibitem[Papantonis and Belle(2025)]%
        {papantonis2025not}
\bibfield{author}{\bibinfo{person}{Ioannis Papantonis} {and} \bibinfo{person}{Vaishak Belle}.} \bibinfo{year}{2025}\natexlab{}.
\newblock \showarticletitle{Why not both? Complementing explanations with uncertainty, and self-confidence in human-AI collaboration}.
\newblock \bibinfo{journal}{\emph{Frontiers in Computer Science}}  \bibinfo{volume}{7} (\bibinfo{year}{2025}), \bibinfo{pages}{1560448}.
\newblock
\href{https://doi.org/10.3389/fcomp.2025.1560448}{doi:\nolinkurl{10.3389/fcomp.2025.1560448}}


\bibitem[Pauli et~al\mbox{.}(2025)]%
        {pauli-etal-2025-measuring}
\bibfield{author}{\bibinfo{person}{Amalie~Brogaard Pauli}, \bibinfo{person}{Isabelle Augenstein}, {and} \bibinfo{person}{Ira Assent}.} \bibinfo{year}{2025}\natexlab{}.
\newblock \showarticletitle{Measuring and Benchmarking Large Language Models' Capabilities to Generate Persuasive Language}. In \bibinfo{booktitle}{\emph{Proceedings of the 2025 Conference of the Nations of the Americas Chapter of the Association for Computational Linguistics: Human Language Technologies (Volume 1: Long Papers)}}, \bibfield{editor}{\bibinfo{person}{Luis Chiruzzo}, \bibinfo{person}{Alan Ritter}, {and} \bibinfo{person}{Lu~Wang}} (Eds.). \bibinfo{publisher}{Association for Computational Linguistics}, \bibinfo{address}{Albuquerque, New Mexico}, \bibinfo{pages}{10056--10075}.
\newblock
\showISBNx{979-8-89176-189-6}
\href{https://doi.org/10.18653/v1/2025.naacl-long.506}{doi:\nolinkurl{10.18653/v1/2025.naacl-long.506}}


\bibitem[Prabhudesai et~al\mbox{.}(2023)]%
        {prabhudesai2023uncertainty}
\bibfield{author}{\bibinfo{person}{Snehal Prabhudesai}, \bibinfo{person}{Leyao Yang}, \bibinfo{person}{Sumit Asthana}, \bibinfo{person}{Xun Huan}, \bibinfo{person}{Q.~Vera Liao}, {and} \bibinfo{person}{Nikola Banovic}.} \bibinfo{year}{2023}\natexlab{}.
\newblock \showarticletitle{Understanding Uncertainty: How Lay Decision-makers Perceive and Interpret Uncertainty in Human-AI Decision Making}. In \bibinfo{booktitle}{\emph{Proceedings of the 28th International Conference on Intelligent User Interfaces}} (Sydney, NSW, Australia) \emph{(\bibinfo{series}{IUI '23})}. \bibinfo{publisher}{Association for Computing Machinery}, \bibinfo{address}{New York, NY, USA}, \bibinfo{pages}{379–396}.
\newblock
\showISBNx{9798400701061}
\href{https://doi.org/10.1145/3581641.3584033}{doi:\nolinkurl{10.1145/3581641.3584033}}


\bibitem[Schemmer et~al\mbox{.}(2023)]%
        {schemmer2023relianceexplanations}
\bibfield{author}{\bibinfo{person}{Max Schemmer}, \bibinfo{person}{Niklas Kuehl}, \bibinfo{person}{Carina Benz}, \bibinfo{person}{Andrea Bartos}, {and} \bibinfo{person}{Gerhard Satzger}.} \bibinfo{year}{2023}\natexlab{}.
\newblock \showarticletitle{Appropriate Reliance on AI Advice: Conceptualization and the Effect of Explanations}. In \bibinfo{booktitle}{\emph{Proceedings of the 28th International Conference on Intelligent User Interfaces}} (Sydney, NSW, Australia) \emph{(\bibinfo{series}{IUI '23})}. \bibinfo{publisher}{Association for Computing Machinery}, \bibinfo{address}{New York, NY, USA}, \bibinfo{pages}{410–422}.
\newblock
\showISBNx{9798400701061}
\href{https://doi.org/10.1145/3581641.3584066}{doi:\nolinkurl{10.1145/3581641.3584066}}


\bibitem[Simon et~al\mbox{.}(2025)]%
        {simon2025generative}
\bibfield{author}{\bibinfo{person}{Felix~M. Simon}, \bibinfo{person}{Rasmus~Kleis Nielsen}, {and} \bibinfo{person}{Richard Fletcher}.} \bibinfo{year}{2025}\natexlab{}.
\newblock \bibinfo{booktitle}{\emph{Generative AI and News Report 2025: How People Think About AI’s Role in Journalism and Society}}.
\newblock \bibinfo{type}{Technical Report}. \bibinfo{institution}{Reuters Institute for the Study of Journalism, University of Oxford}.
\newblock
\href{https://doi.org/10.60625/risj-5bjv-yt69}{doi:\nolinkurl{10.60625/risj-5bjv-yt69}}


\bibitem[Slovic and V{\"a}stfj{\"a}ll(2010)]%
        {slovic2010affect}
\bibfield{author}{\bibinfo{person}{Paul Slovic} {and} \bibinfo{person}{Daniel V{\"a}stfj{\"a}ll}.} \bibinfo{year}{2010}\natexlab{}.
\newblock \showarticletitle{Affect, moral intuition, and risk}.
\newblock \bibinfo{journal}{\emph{Psychological inquiry}} \bibinfo{volume}{21}, \bibinfo{number}{4} (\bibinfo{year}{2010}), \bibinfo{pages}{387--398}.
\newblock
\href{https://doi.org/10.1080/1047840X.2010.521119}{doi:\nolinkurl{10.1080/1047840X.2010.521119}}


\bibitem[Steyvers et~al\mbox{.}(2025)]%
        {steyvers2024calibrationgapmodelhuman}
\bibfield{author}{\bibinfo{person}{Mark Steyvers}, \bibinfo{person}{Heliodoro Tejeda}, \bibinfo{person}{Aakriti Kumar}, \bibinfo{person}{Catarina Belem}, \bibinfo{person}{Sheer Karny}, \bibinfo{person}{Xinyue Hu}, \bibinfo{person}{Lukas~W Mayer}, {and} \bibinfo{person}{Padhraic Smyth}.} \bibinfo{year}{2025}\natexlab{}.
\newblock \showarticletitle{What large language models know and what people think they know}.
\newblock \bibinfo{journal}{\emph{Nature Machine Intelligence}} (\bibinfo{year}{2025}), \bibinfo{pages}{1--11}.
\newblock
\href{https://doi.org/10.1038/s42256-024-00976-7}{doi:\nolinkurl{10.1038/s42256-024-00976-7}}


\bibitem[Sun et~al\mbox{.}(2026)]%
        {sun-etal-2026-explaining}
\bibfield{author}{\bibinfo{person}{Jingyi Sun}, \bibinfo{person}{Greta Warren}, \bibinfo{person}{Irina Shklovski}, {and} \bibinfo{person}{Isabelle Augenstein}.} \bibinfo{year}{2026}\natexlab{}.
\newblock \showarticletitle{Explaining Sources of Uncertainty in Automated Fact-Checking}. In \bibinfo{booktitle}{\emph{Proceedings of the 64th Annual Meeting of the {A}ssociation for {C}omputational {L}inguistics (Volume 1: Long Papers)}}, \bibfield{editor}{\bibinfo{person}{Maria Liakata}, \bibinfo{person}{Viviane~P. Moreira}, \bibinfo{person}{Jiajun Zhang}, {and} \bibinfo{person}{David Jurgens}} (Eds.). \bibinfo{publisher}{Association for Computational Linguistics}, \bibinfo{address}{San Diego, California, United States}, \bibinfo{pages}{45510--45534}.
\newblock
\showISBNx{979-8-89176-390-6}
\href{https://doi.org/10.18653/v1/2026.acl-long.2110}{doi:\nolinkurl{10.18653/v1/2026.acl-long.2110}}


\bibitem[Ulmer et~al\mbox{.}(2026)]%
        {ulmer2026anthropomimetic}
\bibfield{author}{\bibinfo{person}{Dennis Ulmer}, \bibinfo{person}{Alexandra Lorson}, \bibinfo{person}{Ivan Titov}, {and} \bibinfo{person}{Christian Hardmeier}.} \bibinfo{year}{2026}\natexlab{}.
\newblock \showarticletitle{Anthropomimetic uncertainty: What verbalized uncertainty in language models is missing}.
\newblock \bibinfo{journal}{\emph{Transactions of the Association for Computational Linguistics}}  \bibinfo{volume}{14} (\bibinfo{year}{2026}), \bibinfo{pages}{1505--1540}.
\newblock
\href{https://doi.org/10.1162/TACL.a.739}{doi:\nolinkurl{10.1162/TACL.a.739}}


\bibitem[Vasconcelos et~al\mbox{.}(2025)]%
        {Vasconcelos2025generationprob}
\bibfield{author}{\bibinfo{person}{Helena Vasconcelos}, \bibinfo{person}{Gagan Bansal}, \bibinfo{person}{Adam Fourney}, \bibinfo{person}{Q.~Vera Liao}, {and} \bibinfo{person}{Jennifer Wortman~Vaughan}.} \bibinfo{year}{2025}\natexlab{}.
\newblock \showarticletitle{Generation Probabilities Are Not Enough: Uncertainty Highlighting in AI Code Completions}.
\newblock \bibinfo{journal}{\emph{ACM Trans. Comput.-Hum. Interact.}} \bibinfo{volume}{32}, \bibinfo{number}{1}, Article \bibinfo{articleno}{4} (\bibinfo{date}{April} \bibinfo{year}{2025}), \bibinfo{numpages}{30}~pages.
\newblock
\showISSN{1073-0516}
\href{https://doi.org/10.1145/3702320}{doi:\nolinkurl{10.1145/3702320}}


\bibitem[Vasconcelos et~al\mbox{.}(2023)]%
        {vasconcelos_explanations_2023}
\bibfield{author}{\bibinfo{person}{Helena Vasconcelos}, \bibinfo{person}{Matthew Jörke}, \bibinfo{person}{Madeleine Grunde-McLaughlin}, \bibinfo{person}{Tobias Gerstenberg}, \bibinfo{person}{Michael~S. Bernstein}, {and} \bibinfo{person}{Ranjay Krishna}.} \bibinfo{year}{2023}\natexlab{}.
\newblock \showarticletitle{Explanations {Can} {Reduce} {Overreliance} on {AI} {Systems} {During} {Decision}-{Making}}.
\newblock \bibinfo{journal}{\emph{Proceedings of the ACM on Human-Computer Interaction}} \bibinfo{volume}{7}, \bibinfo{number}{CSCW1} (\bibinfo{date}{April} \bibinfo{year}{2023}), \bibinfo{pages}{1--38}.
\newblock
\showISSN{2573-0142}
\href{https://doi.org/10.1145/3579605}{doi:\nolinkurl{10.1145/3579605}}


\bibitem[Wallsten et~al\mbox{.}(1993)]%
        {wallsten1993preferences}
\bibfield{author}{\bibinfo{person}{Thomas~S. Wallsten}, \bibinfo{person}{David~V. Budescu}, \bibinfo{person}{Rami Zwick}, {and} \bibinfo{person}{Steven~M. Kemp}.} \bibinfo{year}{1993}\natexlab{}.
\newblock \showarticletitle{{Preferences and Reasons for Communicating Probabilistic Information in Verbal or Numerical Terms}}.
\newblock \bibinfo{journal}{\emph{Bulletin of the Psychonomic Society}} \bibinfo{volume}{31}, \bibinfo{number}{2} (\bibinfo{year}{1993}), \bibinfo{pages}{135--138}.
\newblock
\href{https://doi.org/10.3758/BF03334162}{doi:\nolinkurl{10.3758/BF03334162}}


\bibitem[Warren et~al\mbox{.}(2025)]%
        {warren2025explainablefactchecking}
\bibfield{author}{\bibinfo{person}{Greta Warren}, \bibinfo{person}{Irina Shklovski}, {and} \bibinfo{person}{Isabelle Augenstein}.} \bibinfo{year}{2025}\natexlab{}.
\newblock \showarticletitle{Show Me the Work: Fact-Checkers' Requirements for Explainable Automated Fact-Checking}. In \bibinfo{booktitle}{\emph{Proceedings of the CHI Conference on Human Factors in Computing Systems}} (Yokohama, Japan) \emph{(\bibinfo{series}{CHI '25})}. \bibinfo{publisher}{Association for Computing Machinery}, \bibinfo{address}{New York, NY, USA}, Article \bibinfo{articleno}{197}, \bibinfo{numpages}{21}~pages.
\newblock
\showISBNx{979-8-4007-1394-1/25/04}
\href{https://doi.org/10.1145/3706598.371327}{doi:\nolinkurl{10.1145/3706598.371327}}


\bibitem[Warren et~al\mbox{.}(2026)]%
        {warren2026evidence}
\bibfield{author}{\bibinfo{person}{Greta Warren}, \bibinfo{person}{Jingyi Sun}, \bibinfo{person}{Irina Shklovski}, {and} \bibinfo{person}{Isabelle Augenstein}.} \bibinfo{year}{2026}\natexlab{}.
\newblock \showarticletitle{Show me the evidence: Evaluating the role of evidence and natural language explanations in AI-supported fact-checking}. In \bibinfo{booktitle}{\emph{Proceedings of the 2026 ACM Conference on Fairness, Accountability, and Transparency}} \emph{(\bibinfo{series}{FAccT '26})}. \bibinfo{publisher}{Association for Computing Machinery}, \bibinfo{address}{New York, NY, USA}, \bibinfo{pages}{4899–4934}.
\newblock
\showISBNx{9798400725968}
\href{https://doi.org/10.1145/3805689.3812358}{doi:\nolinkurl{10.1145/3805689.3812358}}


\bibitem[YouGov(2026)]%
        {yougovMethodology}
\bibfield{author}{\bibinfo{person}{YouGov}.} \bibinfo{year}{2026}\natexlab{}.
\newblock \bibinfo{title}{{M}ethodology --- yougov.com}.
\newblock \bibinfo{howpublished}{\url{https://yougov.com/en-us/about/methodology}}.
\newblock
\newblock
\shownote{[Accessed 24-09-2026]}.


\bibitem[Zhang et~al\mbox{.}(2020)]%
        {zhang2020confidence}
\bibfield{author}{\bibinfo{person}{Yunfeng Zhang}, \bibinfo{person}{Q.~Vera Liao}, {and} \bibinfo{person}{Rachel K.~E. Bellamy}.} \bibinfo{year}{2020}\natexlab{}.
\newblock \showarticletitle{Effect of confidence and explanation on accuracy and trust calibration in AI-assisted decision making}. In \bibinfo{booktitle}{\emph{Proceedings of the 2020 Conference on Fairness, Accountability, and Transparency}} (Barcelona, Spain) \emph{(\bibinfo{series}{FAT* '20})}. \bibinfo{publisher}{Association for Computing Machinery}, \bibinfo{address}{New York, NY, USA}, \bibinfo{pages}{295–305}.
\newblock
\showISBNx{9781450369367}
\href{https://doi.org/10.1145/3351095.3372852}{doi:\nolinkurl{10.1145/3351095.3372852}}


\bibitem[Zimmer(1983)]%
        {zimmer1983verbal}
\bibfield{author}{\bibinfo{person}{Alf~C Zimmer}.} \bibinfo{year}{1983}\natexlab{}.
\newblock \showarticletitle{{Verbal vs. Numerical Processing of Subjective Probabilities}}.
\newblock In \bibinfo{booktitle}{\emph{Advances in psychology}}. Vol.~\bibinfo{volume}{16}. \bibinfo{publisher}{Elsevier}, \bibinfo{pages}{159--182}.
\newblock
\href{https://doi.org/10.1016/S0166-4115(08)62198-6}{doi:\nolinkurl{10.1016/S0166-4115(08)62198-6}}


\end{thebibliography}

\appendix

\section{Observational Survey}
\label{app:survey}

\subsection{Additional Results}
\label{app:rq1:results}

\begin{figure}[h]
    \centering
    \includegraphics[width=\linewidth]{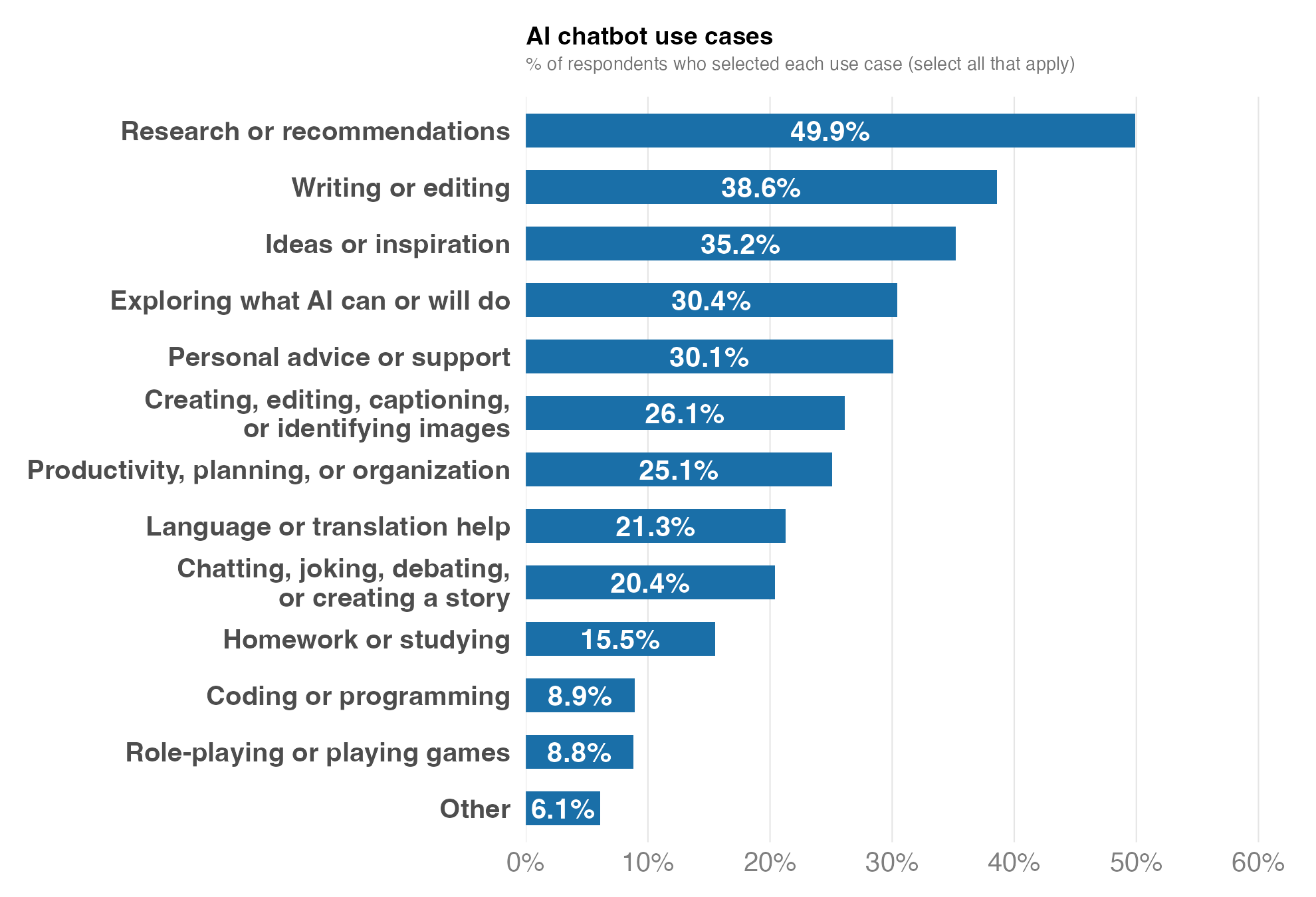}
    \caption{Use cases for AI chatbots}
    \label{fig:app:survey_usecases}
\end{figure}

\subsection{Survey}
Welcome to our survey on generative AI chatbots. We're interested in understanding your experiences and perceptions of these emerging technologies. This survey should take approximately 5 minutes to complete.

Your participation is greatly appreciated, and we value the insights you'll share with us. Your input will help shape the future of AI development, so thank you for taking the time to contribute.

\noindent 1. In what year were you born?

\noindent 2. Are you…?
\begin{itemize}
    \item Male
    \item Female
\end{itemize}

\noindent 3. What is the highest level of education you have completed?

\begin{itemize}
    \item Did not graduate from high school
    \item High school graduate
    \item Some college, but no degree
    \item 2-year college degree
    \item 4-year college degree
    \item Postgraduate degree (MA, MBA, MD, JD, PhD, etc.)
\end{itemize}

\noindent 4. What racial or ethnic group best describes you?

\begin{itemize}
    \item White
    \item Black or African-American
    \item Hispanic or Latino
    \item Asian or Asian-American
    \item Middle Eastern
    \item Two or more races
    \item Other
\end{itemize}

\noindent 5. What is your State of Residence?

[NEXT PAGE]

\noindent 6. How often do you use generative artificial intelligence (AI) chatbots such as the following?

\begin{itemize}
    \item ChatGPT
    \item MetaAI
    \item Gemini
    \item Copilot
    \item Perplexity
    \item Claude
    \item DeepSeek
    \item Grok
    \item Apple Intelligence
\end{itemize}

\noindent Options:

\begin{enumerate}[{(a)}]
\item Never
\item Less than once a month
\item At least once a month
\item At least once a week
\item Every day
\end{enumerate}

\noindent [NEXT PAGE]

\noindent 7. Which of the following, if any, have you used generative artificial intelligence (AI) chatbots for in the past 30 days? Select all that apply.

\begin{enumerate}[{(a)}]
\item Coding or programming
\item Homework or studying
\item Exploring what AI can or will do
\item Chatting, joking, debating, or creating a story
\item Research or recommendations
\item Productivity, planning, or organization
\item Role-playing or playing games
\item Language or translation help
\item Writing or editing
\item Creating, editing, captioning, or identifying images
\item Personal advice or support
\item Ideas or inspiration
\item Other (please specify)
\end{enumerate}

\noindent [NEXT PAGE]

\noindent 8. Thinking about your interactions with generative artificial intelligence (AI) chatbots, please rate the degree to which you find AI-generated responses to be:

Believable

Accurate

Trustworthy

Biased

\begin{enumerate}[{(a)}]
    \item Not at all
    \item Somewhat
    \item Moderately
    \item Quite a bit
    \item Very
\end{enumerate}

\noindent [NEXT PAGE]

\noindent 9. How confident are you in your ability to evaluate the accuracy of the responses you get from generative AI chatbots?

\begin{enumerate}[{(a)}]

\item Not at all confident

\item Somewhat confident

\item Moderately confident

\item Quite confident

\item Extremely confident
\end{enumerate}

\noindent [NEXT PAGE]

\noindent 10. In the past 30 days, how often have you received responses from generative artificial intelligence (AI) chatbots that were false or inaccurate?

\begin{enumerate}[{(a)}]

\item Never

\item Rarely

\item Sometimes

\item Quite often

\item Very often
\end{enumerate}

\noindent [NEXT PAGE]

\noindent 11. When using generative AI chatbots, how often do you take the following steps to verify the accuracy of a response?

\begin{enumerate}[{(a)}]
\item Look up additional information online to confirm the response
\item Review the sources provided by the AI
\item Compare responses from multiple AI chatbots
\item Check what others have said about the topic online
\item Discuss the topic with a friend, colleague, or family member
\item Discuss the topic with someone who has expertise
\item Request a confidence level or rating from the AI
\end{enumerate}

\begin{enumerate}

\item Never

\item Rarely

\item Sometimes

\item Often

\item Always
\end{enumerate}

\noindent [NEXT PAGE]

\noindent  12. Consider a scenario where you're using an AI chatbot to research the topics listed below. For each topic, rate how important it would be for you to know when the AI chatbot is not confident in its response.

\begin{enumerate}[{(a)}]
\item Everyday information needs (e.g., recipe ideas, trip planning)
\item Research tasks such as school or work assignments
\item Financial planning and decision-making
\item Legal matters and advice
\item Health or medical advice
\item Civic engagement and voting information
\item Emotional support and counseling
\item Relationship and family advice
\item Career development guidance
\item Roleplaying, entertainment or games
\end{enumerate}

\begin{enumerate}

\item Not at all important

\item Not that important

\item Moderately important

\item Very important

\item Extremely important
\end{enumerate}

\noindent [NEXT PAGE]

\noindent 13. What motivates you to double-check AI-generated information? (Select all that apply)

\begin{enumerate}[{(a)}]
    \item I want to test how accurate it is
    \item I don't trust AI very much
    \item The subject matter is important or high stakes
    \item The information provided by the AI seems suspect or unreliable
    \item I want to ensure accuracy for my own information
    \item I want to ensure accuracy before I share the information with others
    \item I'm concerned about potential biases in the AI's response
    \item Other, please explain
    \item I don't double-check AI-generated information

\end{enumerate}

\noindent [NEXT PAGE]

\noindent 14. To what extent do you think generative AI chatbots provide sufficient information to help you evaluate the trustworthiness of a response?

\begin{enumerate}[{(a)}]

\item Not at all

\item A little

\item Somewhat

\item Quite a bit

\item Very much
\end{enumerate}

\noindent [NEXT PAGE]

\noindent 15. Please indicate how strongly you agree or disagree with the following statements:
\begin{itemize}
    \item Fact-checking AI responses takes too much time
    \item I'm uncertain about how to verify the accuracy of AI responses
    \item I assume that AI responses are accurate unless proven otherwise
    \item I don't fact-check AI responses unless the information seems obviously false
    \item I often double-check AI responses for accuracy
\end{itemize}

\begin{enumerate}[{(a)}]

\item Strongly disagree

\item Somewhat disagree

\item Neither agree nor disagree

\item Somewhat agree

\item Strongly agree
\end{enumerate}

\noindent [NEXT PAGE]

\noindent Thank you for your participation.

\newpage

\section{Interviews}
\label{app:interviews}
\subsection{Interview Design Stimuli}
\begin{figure*}[h!]
    \centering
    \includegraphics[width=\linewidth]{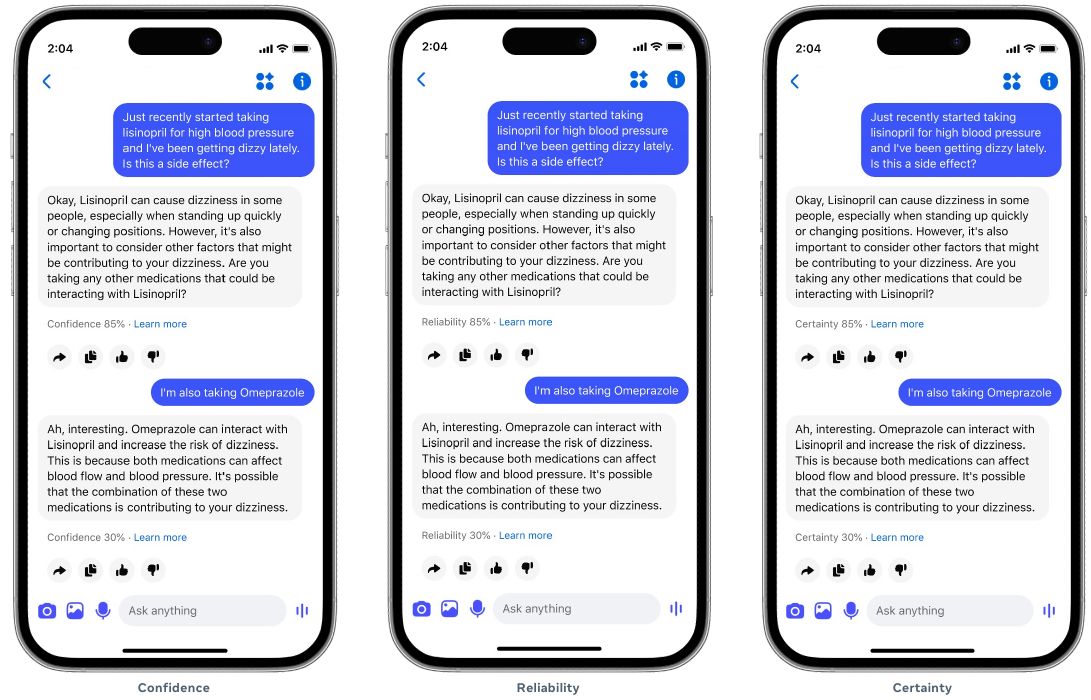}
    \caption{Design stimulus B}
    \label{fig:designstimulusB}
\end{figure*}
\newpage

\begin{figure*}[h!]
    \centering
    \includegraphics[width=\linewidth]{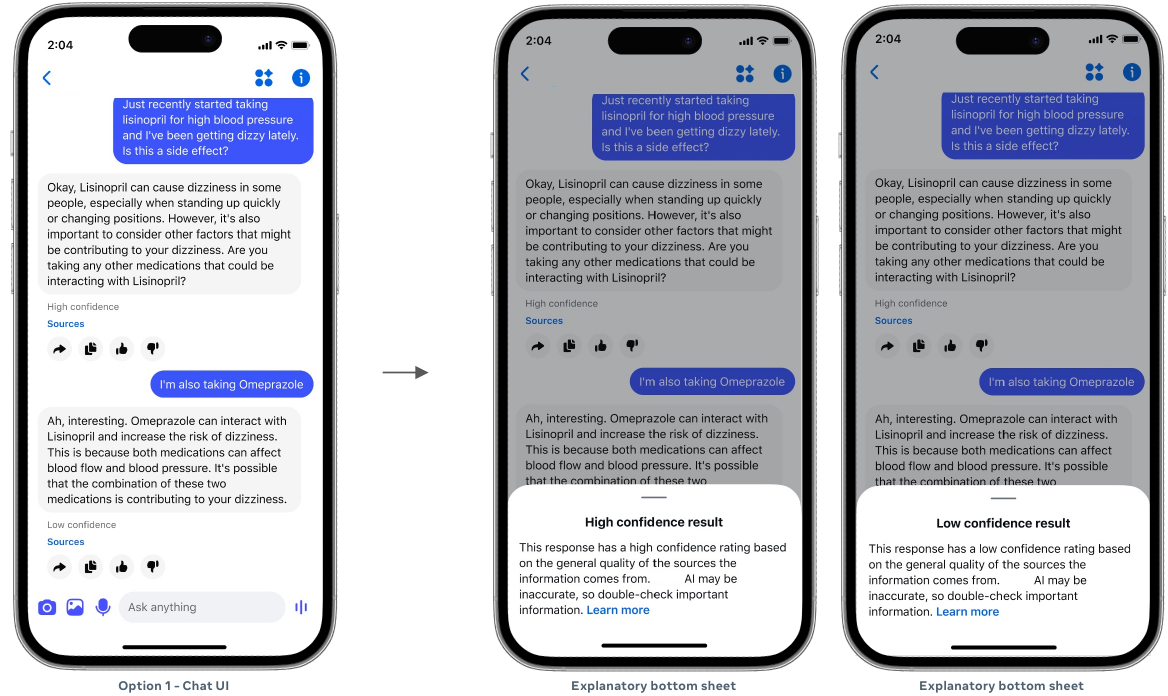}
    \includegraphics[width=\linewidth]{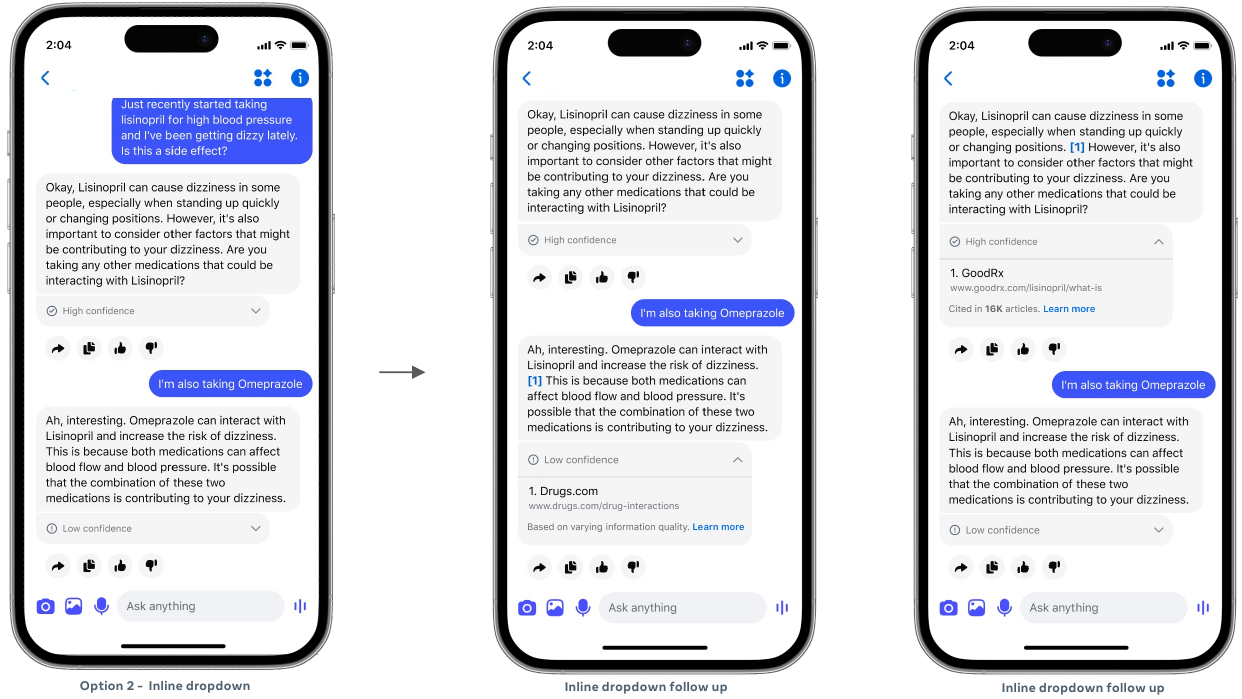}
    \caption{Design stimulus C}
    \label{fig:designstimulusC}
\end{figure*}

\begin{figure*}[h!]
    \centering
    \includegraphics[width=\linewidth]{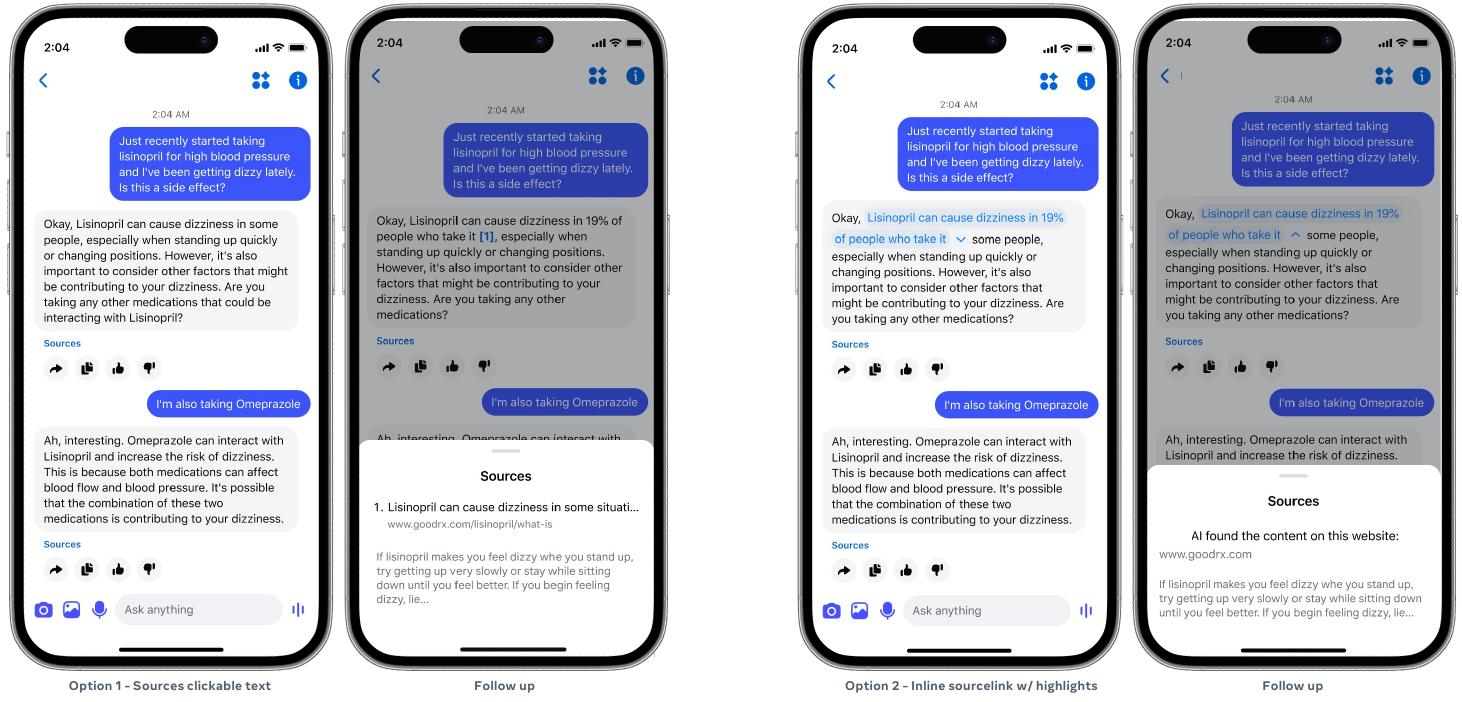}
    \includegraphics[width=.5\linewidth]{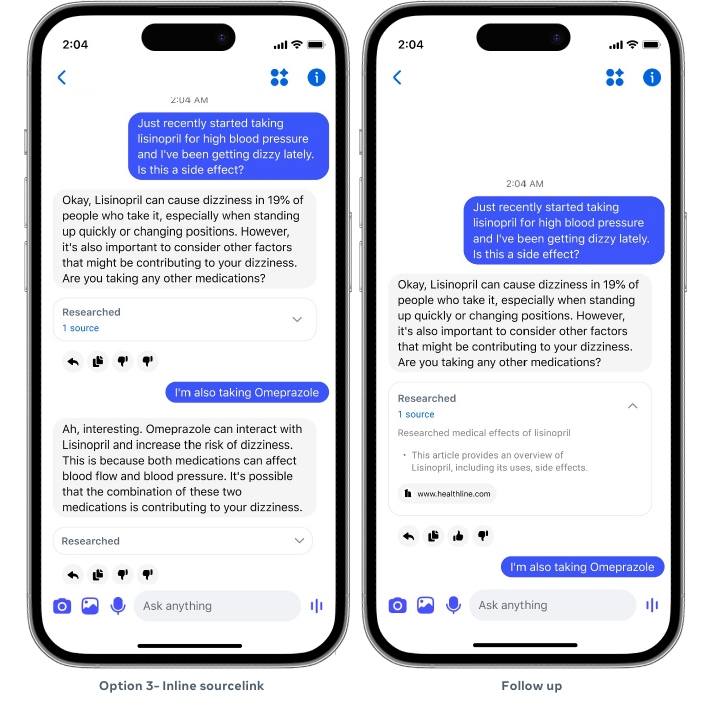}
    \caption{Design stimulus D}
    \label{fig:designstimulusD}
\end{figure*}

\section{Experiment}
\label{app:experiment}

\subsection{Additional Measures}
\label{app:experiment_additional}

\subsubsection{Intent to Use}
There was a main effect of correctness, on participants' self-reported intent to use the chatbot again, F(1, 5115) = 3297.692, p<.001, $\eta^2_p$=.392, indicating that participants were substantially more willing to use chatbots that had provided correct responses ($M = 3.58, SD = 1.15$) than those that had provided incorrect responses ($M = 1.755, SD = 1.132$).
No significant effect of indicator type emerged, F(5, 5115) = 1.034, p=.396, and topic domain did not significantly affect use intentions, F(2, 5115) = 2.825 , p=.059, suggesting that willingness to use the chatbot after accuracy revelation was determined almost entirely by response correctness.

\subsubsection{Recommendability}
There was a main effect of correctness on participants intent to recommend the AI chatbot to others, F(1, 5115) = 2829.25, p<.001, $\eta^2_p$=.356, indicating that participants were substantially more likely to recommend chatbots that had provided correct responses ($M = 3.32, SD = 1.26$) than those that had provided incorrect responses ($M = 1.6, SD = 1.05$).
There was also a main effect of topic, F(2, 5115) = 4.31 , p=.014, $\eta^2_p$=.002. Post hoc Tukey HSD tests indicated that participants in the Gardening condition ($M = 2.50, SD = 1.44$) were significantly more likely to recommend the chatbot than those in the Medical condition ($M = 2.38, SD =1.42$), p = .011 , d = .07. No differences emerged between participants in the Math condition ($M = 2.46, SD = 1.47$) and the others, both ps >.05.
There was no effect of factuality indicator, F(5, 5115) = 1.09, p=.364 on people's likelihood of recommending the chatbot to others.

\subsection{Experiment Items}
\label{app:experiment_items}
Welcome to our survey on generative AI chatbots. We're interested in understanding your perceptions of these emerging technologies. We have tested the survey and found that on average it takes around 10 minutes to complete.

Your participation is greatly appreciated, and we value the insights you'll share with us. Your input will help shape the future of AI development, so thank you for taking the time to contribute.

Please click the forward button below to continue.

\noindent 1. In what year were you born?

\noindent 2. Are you…?
\begin{itemize}
    \item Male
    \item Female
\end{itemize}

\noindent 3.
What is the highest level of education you have completed?

\begin{enumerate}
    \item Did not graduate from high school
    \item High school graduate
    \item Some college, but no degree (yet)
    \item 2-year college degree
    \item 4-year college degree
    \item Postgraduate degree (MA, MBA, MD, JD, PhD, etc.)
\end{enumerate}

\noindent 4. What racial or ethnic group best describes you?

\begin{enumerate}
    \item White
    \item Black or African-American
    \item Hispanic or Latino
    \item Asian or Asian-American
    \item Native American
    \item Middle Eastern
    \item Two or more races
    \item Other
\end{enumerate}

\noindent 5. What is your State of Residence?

[NEXT]

How often do you use generative artificial intelligence (AI) chatbots such as the following?

\begin{itemize}
    \item ChatGPT
    \item MetaAI
    \item Gemini
    \item Copilot
    \item Perplexity
    \item Claude
    \item DeepSeek
    \item Grok
    \item Apple Intelligence
\end{itemize}

 \begin{enumerate}[(a)]
     \item Never
     \item Less than once a month
     \item At least once a month
     \item At least once a week
     \item Every day
 \end{enumerate}

[NEXT]

\noindent Below is an example of a conversation with a generative AI chatbot. Please carefully review the image(s) and respond to the questions based on the content you see. You will only be able to view this once and must review it for at least 30 seconds before moving forward.

[DESIGN SHOWN]

\noindent 1. Overall, how would you describe your impression of this AI chatbot? [Very positive, Slightly positive, Neither positive nor negative, Slightly negative - Very negative]

\noindent 2. How much would you trust the responses from this AI chatbot? [Very much - Quite a bit - Somewhat - A little - Not at all]

\noindent 3. To what extent does this AI chatbot provide sufficient information to help you evaluate the accuracy of the response? [Very much - Quite a bit - Somewhat - A little - Not at all]

\noindent 4. How accurate or inaccurate do you think the response from this AI chatbot is?

<1> Completely inaccurate

<2> Mostly inaccurate

<3> Partially accurate

<4> Mostly accurate

<5> Completely accurate

[NEXT]

[ACCURACY REVELATION]

[ACCURATE] The correct response to the question “ABC” is “XYZ”, meaning, the AI chatbot was correct.

OR

[INACCURATE] The correct response to the question “ABC” is “XYZ”, meaning, the AI chatbot was incorrect.

[NEXT]

\noindent Knowing that the response was correct, please rate your final impression of the AI chatbot.

OR

\noindent Knowing that the response was incorrect, please rate your final impression of the AI chatbot.

\noindent 5. Overall, how would you describe your impression of this AI chatbot? [Very positive, Slightly positive, Neither positive nor negative, Slightly negative - Very negative]

\noindent 6. How much would you trust the responses from this AI chatbot? [Very much - Quite a bit - Somewhat - A little - Not at all]

\noindent 7. How likely is it that you would use this AI chatbot?

<1> Not at all likely

<2> Somewhat likely

<3> Moderately likely

<4> Very likely

<5> Extremely likely

\noindent 8. How likely is it that you would recommend this AI chatbot to others?

<1> Not at all likely

<2> Somewhat likely

<3> Moderately likely

<4> Very likely

<5> Extremely likely

\end{document}